\documentclass[draft]{agujournal2019}
\journalname{XXX}
\usepackage{url}
\usepackage{soul}
\usepackage{caption}
\usepackage{subcaption}
\usepackage{graphicx}
\usepackage{natbib}
\usepackage{amsmath}
\usepackage{bm}
\usepackage{amssymb}

\def\ee{\mathrm{e}}
\def\ii{\mathrm{i}}
\newcommand{\diff}{\, \mathrm{d}}
\newcommand{\LL}{\left}
\newcommand{\RR}{\right}

\newcommand{\hwvec}[1]{\bm{#1}}
\newcommand{\beq}{\begin{equation}}
\newcommand{\eeq}{\end{equation}}
\newcommand{\cosb}[1]{\cos{\LL(#1\RR)}}
\newcommand{\sinb}[1]{\sin{\LL(#1\RR)}}

\newcommand{\ave}[1]{\left<{#1}\right>}
\newcommand{\avesq}[1]{\left<{\LL(#1\RR)}^2\right>}

\newcommand{\xs}{x_{\mathrm{s}}} 
\newcommand{\xo}{x_{\mathrm{o}}} 
 
\newcommand{\rs}{r_{\mathrm{s}}} 
\newcommand{\ro}{r_{\mathrm{o}}} 
\newcommand{\ers}{\hwvec{e_{r\mathrm{s}}}} 
\newcommand{\ero}{\hwvec{e_{r\mathrm{o}}}} 
\newcommand{\ez}{\hwvec{e_{z}}}

\newcommand{\alphao}{\alpha_{\mathrm{o}}}

\newcommand{\hwdef}{{\stackrel{\,\text{def}}{=}}\,\, }

\newcommand{\duo}{\Delta u_{\mathrm{o}}} 
\newcommand{\dus}{\Delta u_{\mathrm{s}}}

\newcommand{\duls}{\Delta u_{\mathrm{Ls}}} 
\newcommand{\dulf}{\Delta u_{\mathrm{Lf}}}  
\newcommand{\dulo}{\Delta u_{\mathrm{Lo}}}  
\newcommand{\duts}{\Delta u_{\mathrm{Ts}}}
\newcommand{\dutf}{\Delta u_{\mathrm{Tf}}} 
\newcommand{\duto}{\Delta u_{\mathrm{To}}}

\newcommand{\DLO}{D_{\mathrm{LLo}}}
\newcommand{\DTO}{D_{\mathrm{TTo}}}
\newcommand{\DDO}{D_{\mathrm{\phi\phi o}}}
\newcommand{\DRO}{D_{\mathrm{\psi\psi o}}}
\newcommand{\DLS}{D_{\mathrm{LLs}}}
\newcommand{\DTS}{D_{\mathrm{TTs}}}
\newcommand{\DDS}{D_{\mathrm{\phi\phi s}}}
\newcommand{\DRS}{D_{\mathrm{\psi\psi s}}}
\newcommand{\DLF}{D_{\mathrm{LLf}}}
\newcommand{\DTF}{D_{\mathrm{TTf}}}
\newcommand{\DDF}{D_{\mathrm{\phi\phi f}}}
\newcommand{\DRF}{D_{\mathrm{\psi\psi f}}}
\newcommand{\DLi}{D_{\mathrm{LLi}}}
\newcommand{\DTi}{D_{\mathrm{TTi}}}
\newcommand{\DDi}{D_{\mathrm{\phi\phi i}}}
\newcommand{\DRi}{D_{\mathrm{\psi\psi i}}}
\newcommand{\rj}{r_{\mathrm{j}}}

\newcommand{\DXXxn}[2]{D^{#2}_{\mathrm{#1}}}

\definecolor{HW}{RGB}{137,0,225}
\newcommand{\HW}[1]{#1}
\newcommand{\HWn}[1]{#1}

\renewcommand{\thesubsection}{\thesection\alph{subsection}}

\begin{document}
  
\title{Wave-mean decomposition of scale-dependent kinetic energy from surface drifters}

\authors{Han Wang\affil{1}, Dhruv Balwada\affil{2}, and Jin-Han Xie\affil{3}}

\affiliation{1}{Institute of Oceanography, University of Hamburg, 20146 Hamburg, Germany}
\affiliation{2}{Lamont Doherty Earth Observatory, Columbia University, Palisades, NY, USA}
\affiliation{3}{School of Mechanics and Engineering Science, Peking University, China}

\correspondingauthor{Han Wang}{hannnwangus@gmail.com}
%
%

\begin{abstract}
Separating waves and mean flows is a fundamental challenge in ocean dynamics.
Lagrangian filtering of passive-tracer time series into high-frequency wave and low-frequency mean-flow components provides a practical route, as the relevant time scales are often cleanly split in the Lagrangian frame. 
Here we show that Lagrangian filtering can be applied to surface drifter observations, providing a powerful approach to quantify wave and mean-flow contributions to surface kinetic energy statistics. 
A key methodological choice is to implement the filtering in a generalized Lagrangian mean (GLM) framework, attributing filtered velocities to mean rather than particle trajectories; this produces more physically interpretable diagnostics. 
Using Gulf of Mexico drifter data, we compute second-order velocity structure functions (SF2s) for waves and mean flow components across spatial scales.  With these filtered SF2s as a benchmark, we illustrate that Helmholtz decomposition of unfiltered SF2s alone should not be interpreted as a dynamical wave–mean decomposition. 
Applying Helmholtz decomposition to the filtered SF2s further illuminates seasonal dynamics. 
Mean-flow surface kinetic energy is rotationally dominated at scales larger than $O(1)$~km, while at and below $O(1)$ km, divergent and rotational contributions are approximately equipartitioned in both summer and winter, suggesting low-frequency divergent motions and possible associated vertical exchange.
Winter mean flows are more active than summer mean flows over 500 m–10 km.
Super-inertial motions are broadly consistent with linear waves. 
In winter, wave kinetic energy is concentrated at smaller spatial scales than in summer, possibly reflecting enhanced downscale transfer by stronger submesoscale mean flows.
\end{abstract}

\section{Introduction}
Kinetic energy in the ocean is shared between balanced or slow motions  — such as mesoscale eddies and submesoscale currents, whose velocity fields are constrained by diagnostic balance relations — and unbalanced or fast motions, predominantly internal gravity waves, which propagate through oscillatory dynamics. 
Understanding how energy is partitioned between these ``balanced'' and ``unbalanced'' regimes across scales is central for diagnosing 
oceanic energy pathways \citep{2009_ARFM_FerrariWunsch, alford2007internal}, understanding oceanic transport \citep{balwada2018submesoscale}, and  interpreting remote-sensing and in-situ observations \citep{klein2019ocean}.
The last point is particularly relevant for observations from the new generation of satellites such as the Surface Water and Ocean Topography (SWOT) mission, which samples the scales where balanced motions and unbalanced motions co-exist. 

Achieving the balanced-unbalanced decomposition from observations remains challenging.
A widely used diagnostic is the Helmholtz decomposition, which separates horizontal velocities, or statistics derived from them, into rotational and divergent components. 
While observational data to perform this decomposition directly on velocities are rarely available \citep{farrar2020s}, as doing so needs spatially two-dimensional velocity measurements, it is possible to obtain a decomposition of statistical metrics averaged over repeated measurements even when they are constrained in spatial coverage. 
Applied to scale-dependent statistics such as second-order structure functions (SF2s) or wavenumber spectra, the Helmholtz decomposition has been used on ship-track, aircraft, surface-drifters, and model data \citep{buhler2014wave, lindborg2015helmholtz, balwada2016scale, balwada2022direct, sinha2019modulation}. 
As a rough heuristic, rotational velocities are often associated with balanced flows, whereas divergent velocities are often associated with unbalanced flows. This heuristic is motivated by the fact that geostrophic and low-Rossby-number balanced flows are one of the most energetic components of balanced flows \citep{2009_ARFM_FerrariWunsch} and are dominated by rotational motions, while super-inertial internal waves are one of the most energetic components of unbalanced flows \citep{alford2007internal} and possess strong divergent motions. 


This heuristic is not an accurate dynamical decomposition: balanced flows can be substantially divergent (e.g., submesoscale fronts statistically have more rotational imprints than divergent imprints \citep{barkan2019role}), and unbalanced flows can be substantially rotational (e.g., near-inertial waves). 
Dynamical extensions of the Helmholtz decomposition can be based on using buoyancy data and lead to a decomposition into  linear waves and geostrophic \citep{buhler2014wave} or weakly nonlinear balanced flows \citep{wang2020ageostrophic}.
In practice, however, concurrent buoyancy measurements at the required resolution are often unavailable.
\citet{buhler2014wave} suggested that when buoyancy measurements are unavailable, the Garrett–Munk internal wave spectrum \citep{garrett1972space} can be used to model the potential energy contribution, though this does not account for regional or seasonal deviations from the Garrett-Munk spectrum. 
Recent works relax this data requirement by using other types of additional information: \citet{vladoiu2024energy} applied a full wave–mean decomposition to shipboard ADCP and towed CTD data in the pycnocline, and \citet{vladoiu2026method} exploited the distinct polarization relations of waves and vortical modes to achieve a wave-mean decomposition using depth-resolved velocity measurements. 

Other decomposition approaches require different types of observational or modeling information. Eulerian temporal filtering can separate frequency bands from moored or fixed-point measurements, so that fast motions may be treated as unbalanced motions and slow motions as balanced motions. However, Doppler shifting by the balanced motions smears the spectral gap between balanced and unbalanced motions, making clean separation difficult in energetic regions \citep{gerkema2013note,pinkel2008advection}. Vertical-mode methods require resolved velocity profiles and assumptions about the vertical modal structures \citep{torres2022separating}.
In three-dimensional gridded simulations, the flow can be projected onto the linear wave and geostrophic modes \citep{bartello1995geostrophic, Early2021},
with nonlinear extensions available in idealized settings \citep{viudez2004optimal,eden2019gravity, chouksey2018internal}.

Lagrangian filtering offers an alternative approach that circumvents many of the limitations or assumptions described above. 
It exploits the temporal separation between balanced and unbalanced motions by applying frequency filters to time series collected along flow-following trajectories.
The key advantage of working in the Lagrangian frame is that the balanced-motions' Doppler shift, which challenges Eulerian filtering \citep{gerkema2013note, pinkel2008advection}, is largely removed. As a result, the time-scale separation between balanced and unbalanced motions is typically cleaner in the Lagrangian frame than in the Eulerian frame, even after accounting for fast-evolving balanced motions such as submesoscale fronts \citep{callies2020time,jones2023using}. Lagrangian filtering has been developed and tested using synthetic particles in ocean models, and has been shown to outperform Eulerian frequency filtering in preserving transport-relevant balanced motions, including convergent flows near fronts \citep{shakespeare2021new, jones2023using}. Grid-based implementations that avoid explicit particle tracking have also been developed \citep{kafiabad2022grid, baker2025lagrangian}. 



Surface drifters (``drifters'') provide flow-following time series directly, making them a natural platform for Lagrangian filtering. 
However, drifter observations are spatially sparse and unstructured, so they cannot be used to directly reconstruct spatially resolved flow fields. What drifters do provide is access to scale-dependent statistics: SF2s computed from velocity differences between pairs of drifters as a function of their separation distance characterize kinetic energy across spatial scales \citep{bennett1984relative, davidson2015turbulence, balwada2016scale}. Combined with  Helmholtz decomposition, SF2s can further distinguish rotational and divergent contributions \citep{lindborg2015helmholtz, balwada2016scale}. 
Previous drifter SF2 analyses, however, have mostly (with the only exception of \citet{beron2016statistics} to our knowledge)  used unfiltered velocities, and therefore do not decompose balanced and unbalanced contributions.
Combining Lagrangian filtering with SF2s offers a way to quantify how  balanced and unbalanced motions each contribute to surface kinetic energy across scales.

Here we apply Lagrangian frequency filtering to drifter data from the Grand Lagrangian Deployment (GLAD; summer 2012, \citet{poje2014submesoscale}) and the Lagrangian Submesoscale Experiment (LASER; winter 2016, \citet{dasaro2018ocean}), both conducted in the Gulf of Mexico.
\HWn{We apply a frequency filter with cutoff frequency near the inertial frequency to obtain lowpass and highpass velocity components, which we refer to as ``mean flows" and ``waves", respectively, following standard wave-mean terminology \citep{buhler2014b}. 
Due to the time scale separation between balanced and unbalanced motions in the Lagrangian frame \citep{callies2020time,jones2023using}, we interpret the waves (highpass) and mean (lowpass) flows  as unbalanced and balanced motions respectively.
} 
We compute SF2s for both components, and apply the Helmholtz decomposition to these filtered SF2s.
A key methodological choice is to implement the filtering within the generalized Lagrangian-mean (GLM) framework \citep{andrews1978exact, buhler2014b}, attributing both highpass and lowpass velocities to the lowpass-filtered trajectory rather than to the unaveraged particle trajectory. This yields, to our knowledge, the first GLM-inspired wave–mean decomposition of scale-dependent kinetic energy from ocean observations. 
We use the results both methodologically, to compare the GLM-inspired decomposition with other decomposition approaches, and dynamically, to examine seasonally varying wave and mean-flow contributions to surface kinetic energy in the Gulf of Mexico.

The contributions of different frequency components in SF2s have been studied in several contexts.  \cite{beron2016statistics}, which came closest to the analyses here, compared SF2s before and after inertial oscillations are removed by Lagrangian frequency filtering. However, this study focused on pair dispersion, rather than doing a detailed exploration of the filtering methodology or all the insights that can be derived from such an approach.  
\citet{callies2020time} proposed a frequency-resolving SF2 for Eulerian data; we are not aware of its clear connections to Lagrangian data. \citet{hypolite2026second} analyzed SF2s before and after Eulerian frequency filtering in simulations. Our approach differs in using GLM-inspired Lagrangian filtering as the starting point. 

The paper is organized as follows. Section 2 describes the data and methods. Section 3 discusses methodological lessons on wave-mean decomposition. Section 4 presents dynamical insights for the Gulf of Mexico. Section 5 summarizes the results and discusses directions for future work.


\section{Data and Methods}
\subsection{GLAD and LASER datasets} \label{sec:GLADLASER}
\begin{figure}
         \centering
                  \includegraphics[width=0.49\textwidth]
         {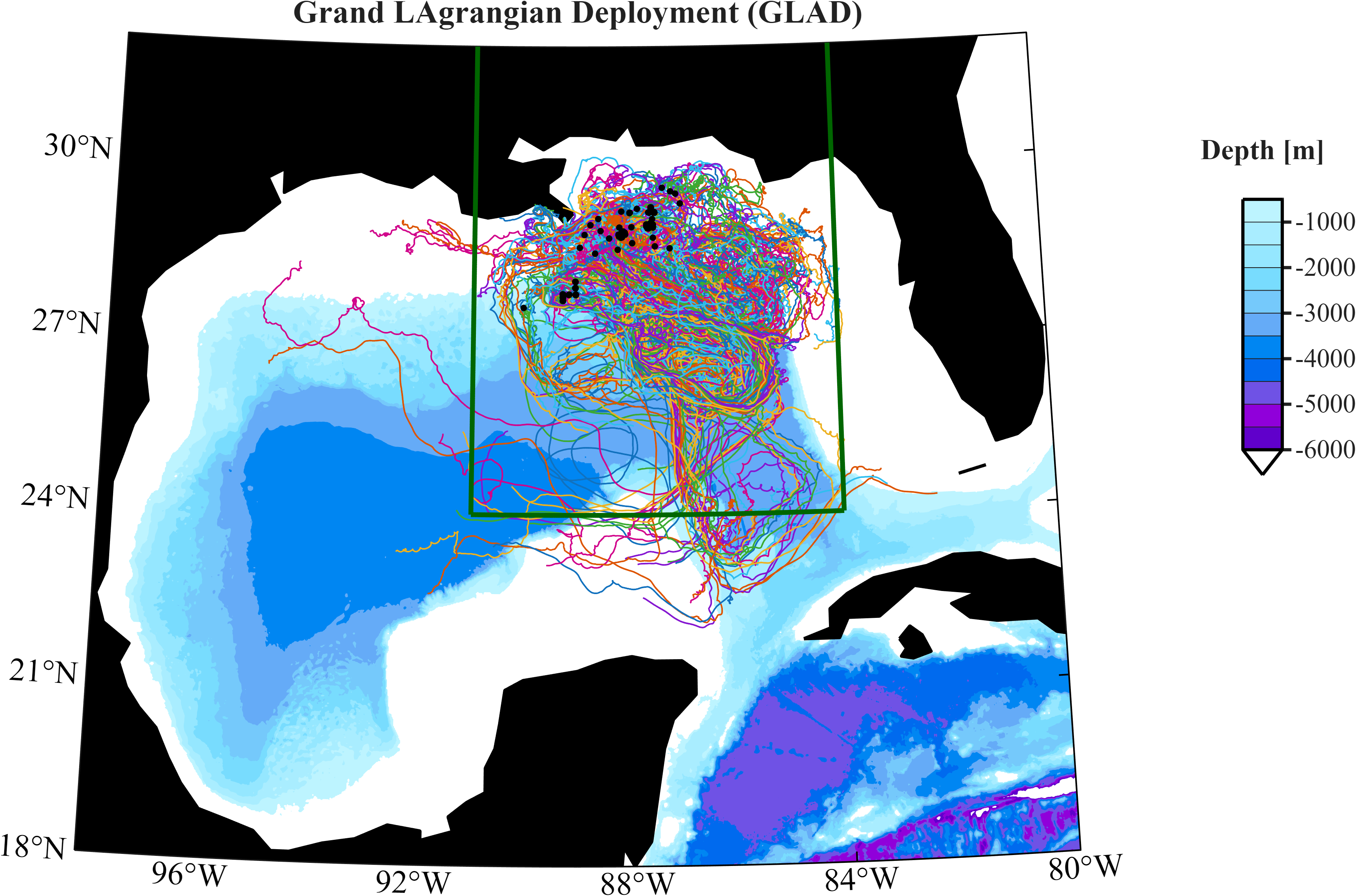}
         \includegraphics[width=0.49\textwidth]{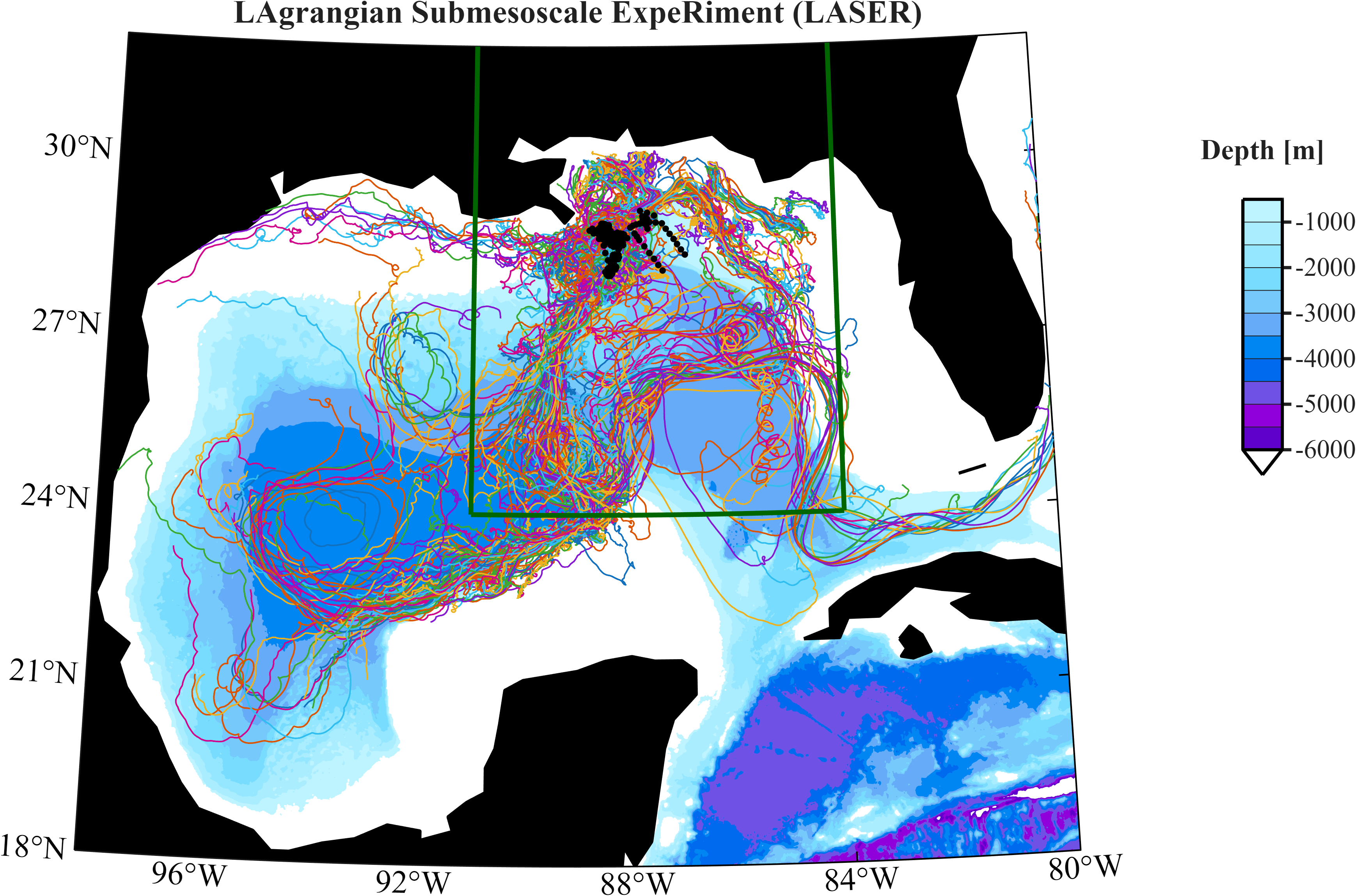}
         \includegraphics[width=0.45\textwidth]
         {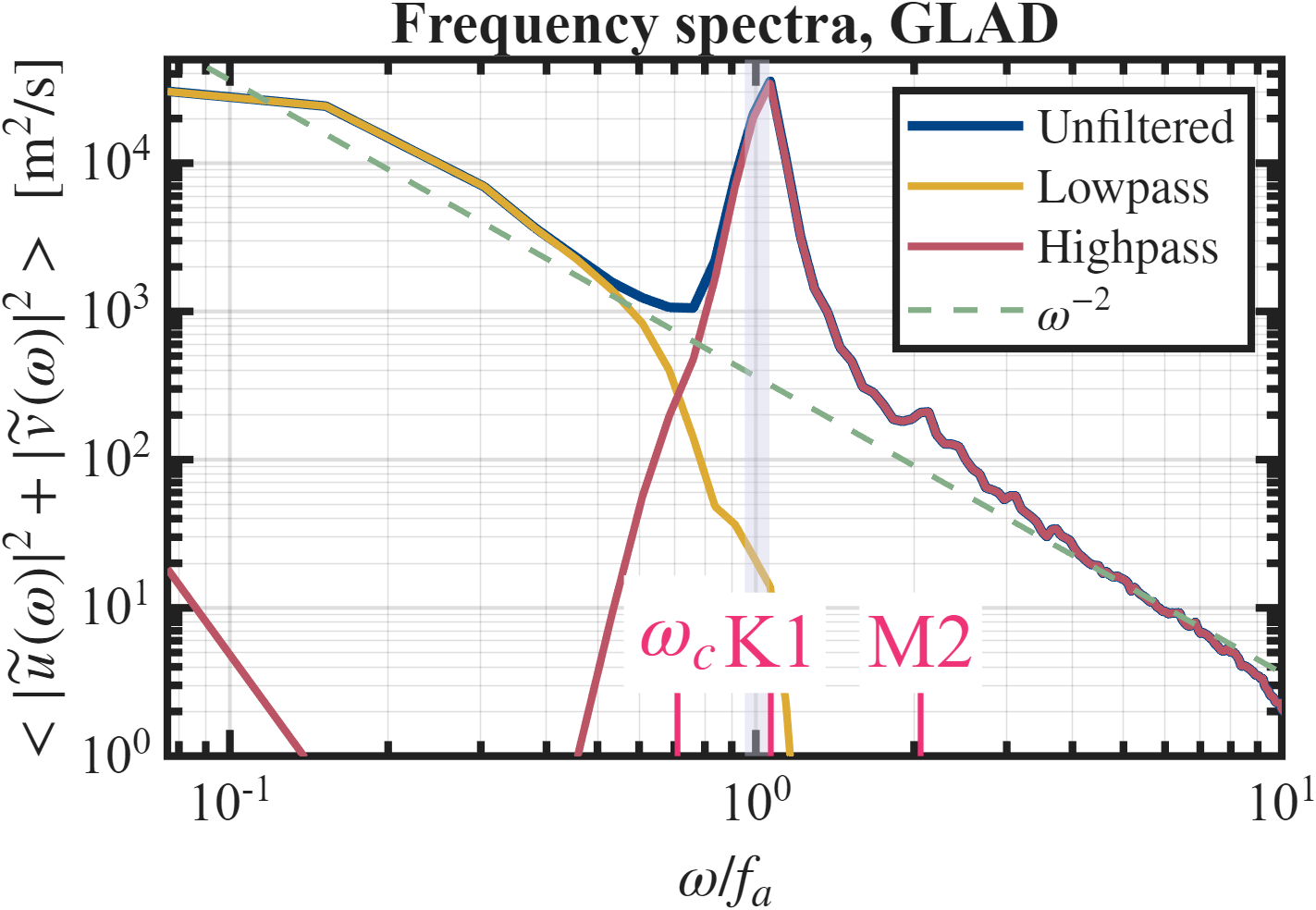}
         \includegraphics[width=0.45\textwidth]{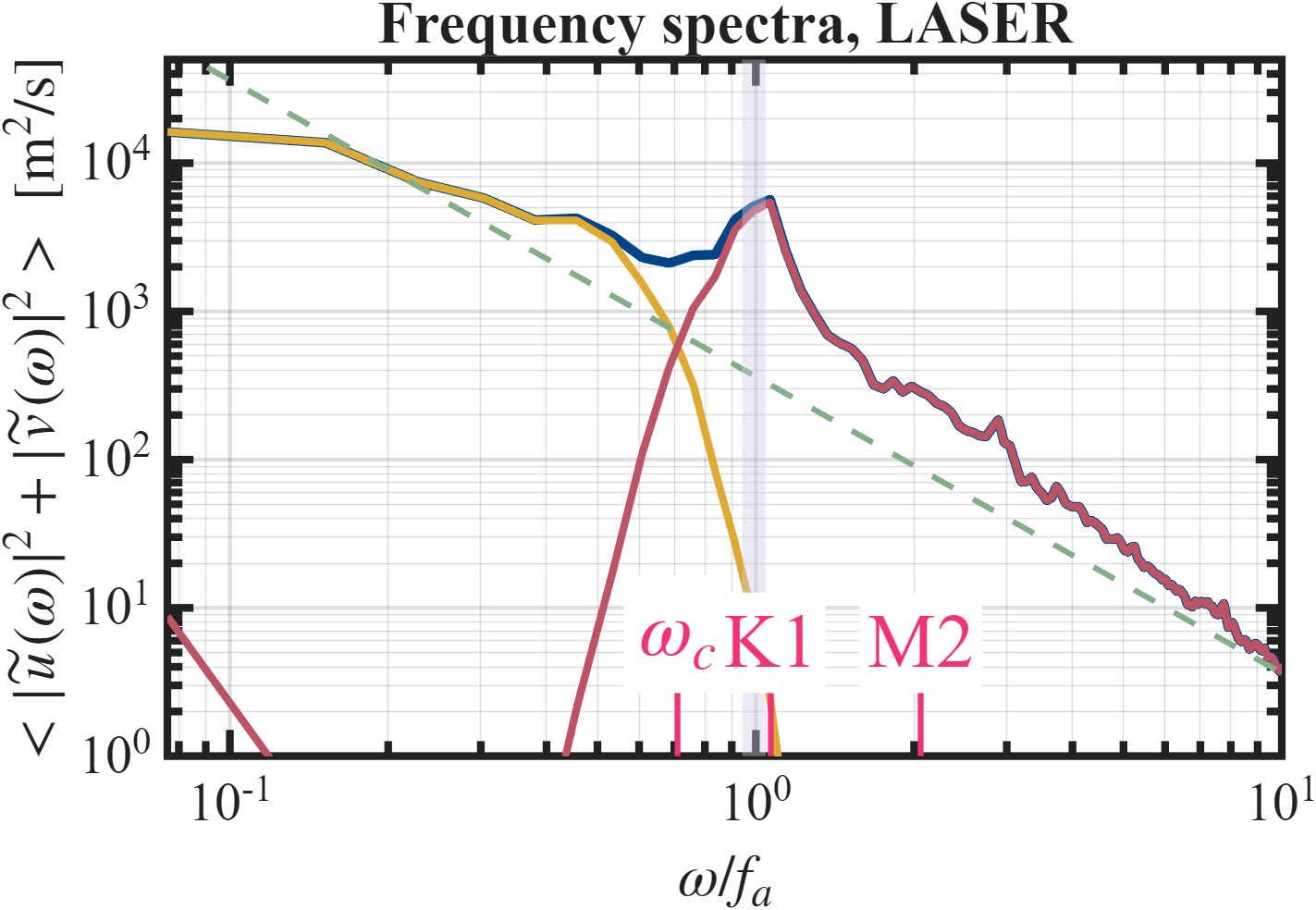}
     \caption{Upper row: all trajectories in the GLAD (left) and LASER campaigns (right) from data products \citep{GLAD15min,LASER15min}. Black dots indicate the drifters' starting positions.     
     Colored contours show bottom bathymetry; regions shallower than $500$~m are left blank. Green boxes mark the range of latitudes and longitudes used in the analysis. 
     Lower row: frequency spectra before (``Unfiltered'') and after (``lowpass'' and ``highpass'') Lagrangian frequency filtering in GLAD (left) and LASER (right). Horizontal axis is scaled by the typical Coriolis frequency $f_a$ evaluated at $28^\circ$ N. The range of Coriolis frequencies covered by all the data points included in the SF2 analysis is indicated by the shading around $\omega/f_a=1$. The cutoff frequency of the frequency filter, corresponding to a period of 1.5 days, is denoted by $\omega_c$ on the horizontal axis. The diurnal and semidiurnal tidal frequencies are denoted by K1 and M2 on the horizontal axis. Dashed lines mark power laws $\propto \omega^{-2}$, consistent with Ornstein–Uhlenbeck process and often found in other oceanic Lagrangian kinetic energy measurements \citep{d2000lagrangian, lacasce2008statistics}.
     Values below $1$~m$^2/$s  are omitted. }\label{fig:GLADLASERoverview}  
\end{figure}

We apply our methods to drifter observations from two observational campaigns in the Gulf of Mexico: the Grand Lagrangian Deployment (GLAD) in summer 2012 (20 July--22 October 2012) and the Lagrangian Submesoscale Experiment (LASER) in winter 2016 (21 January--30 April 2016). The spatial extent of the drifter trajectories from these campaigns is shown in the upper row of Fig.~\ref{fig:GLADLASERoverview}. 

\begin{figure}
  \centering
  \begin{minipage}[t]{0.55\textwidth}
    \centering
    \vspace{0pt}
    \setlength{\parskip}{0pt}%
    \includegraphics[width=0.9\textwidth]{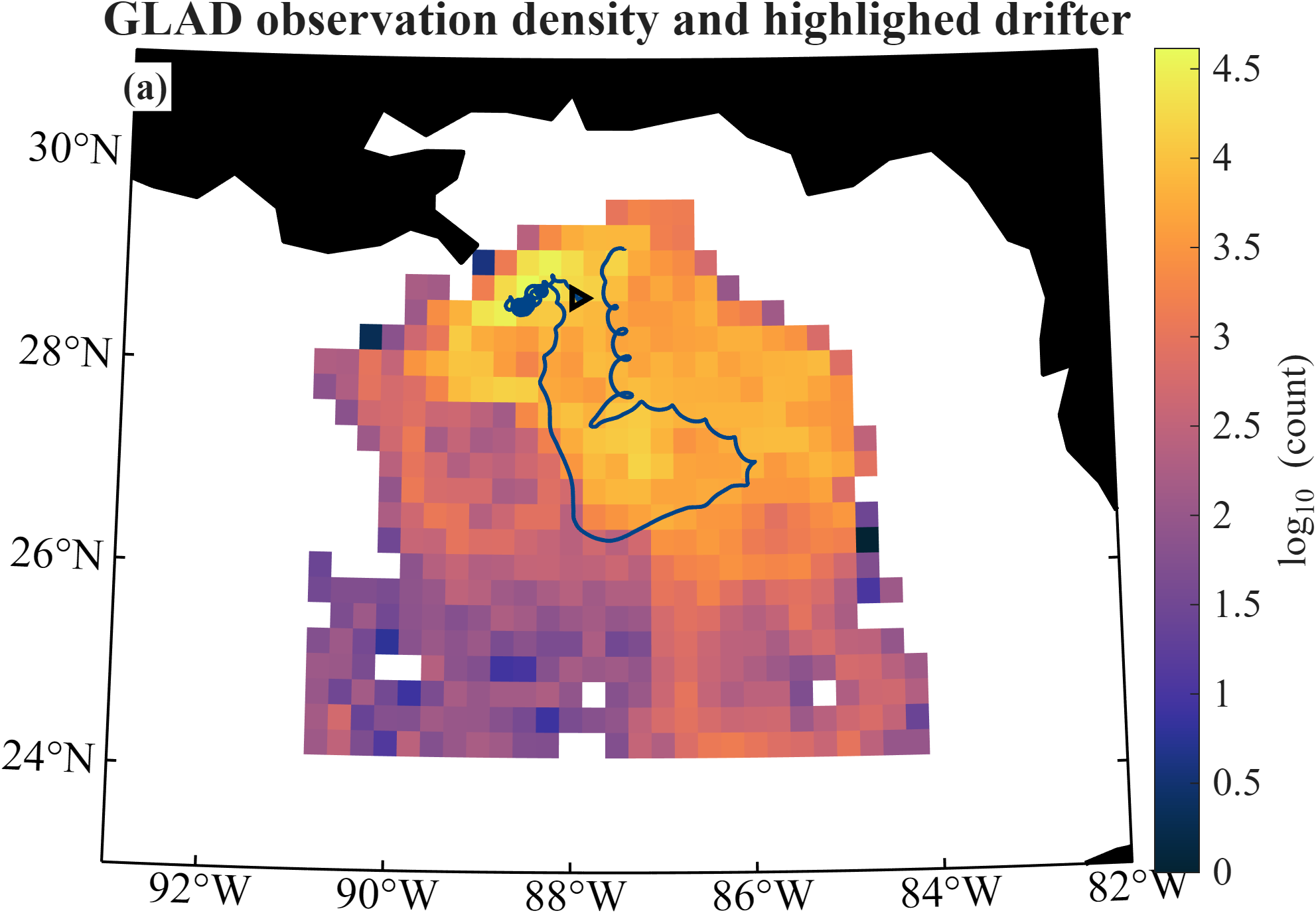}
    \vspace{1.6em}
    \includegraphics[width=\textwidth]{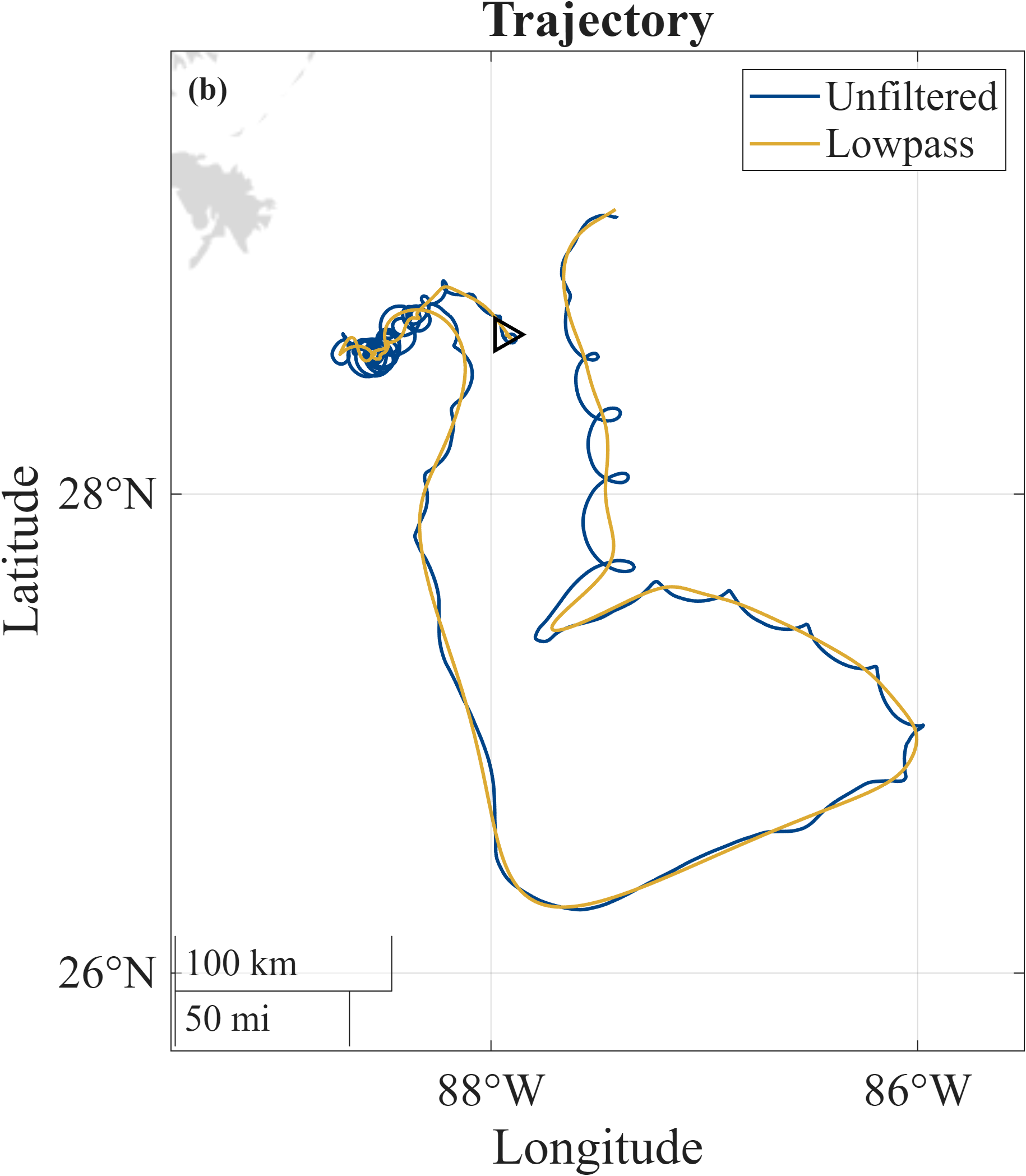}
  \end{minipage}%
  \hfill
  \begin{minipage}[t]{0.44\textwidth}
    \centering
    \vspace{0pt}
    \setlength{\parskip}{0pt}
    \includegraphics[width=\textwidth]{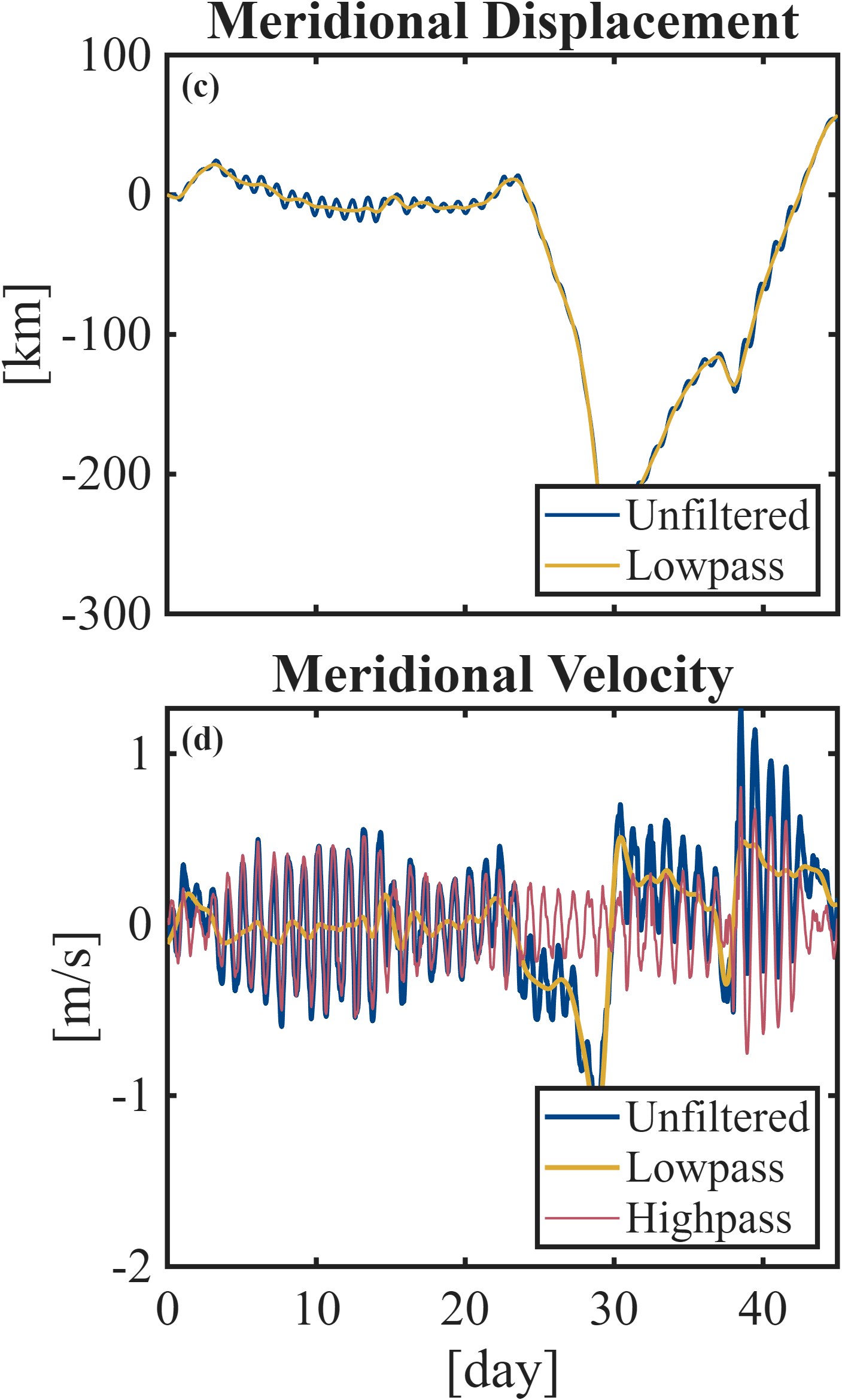}
    \vspace{0.6em}
    \includegraphics[width=\textwidth]{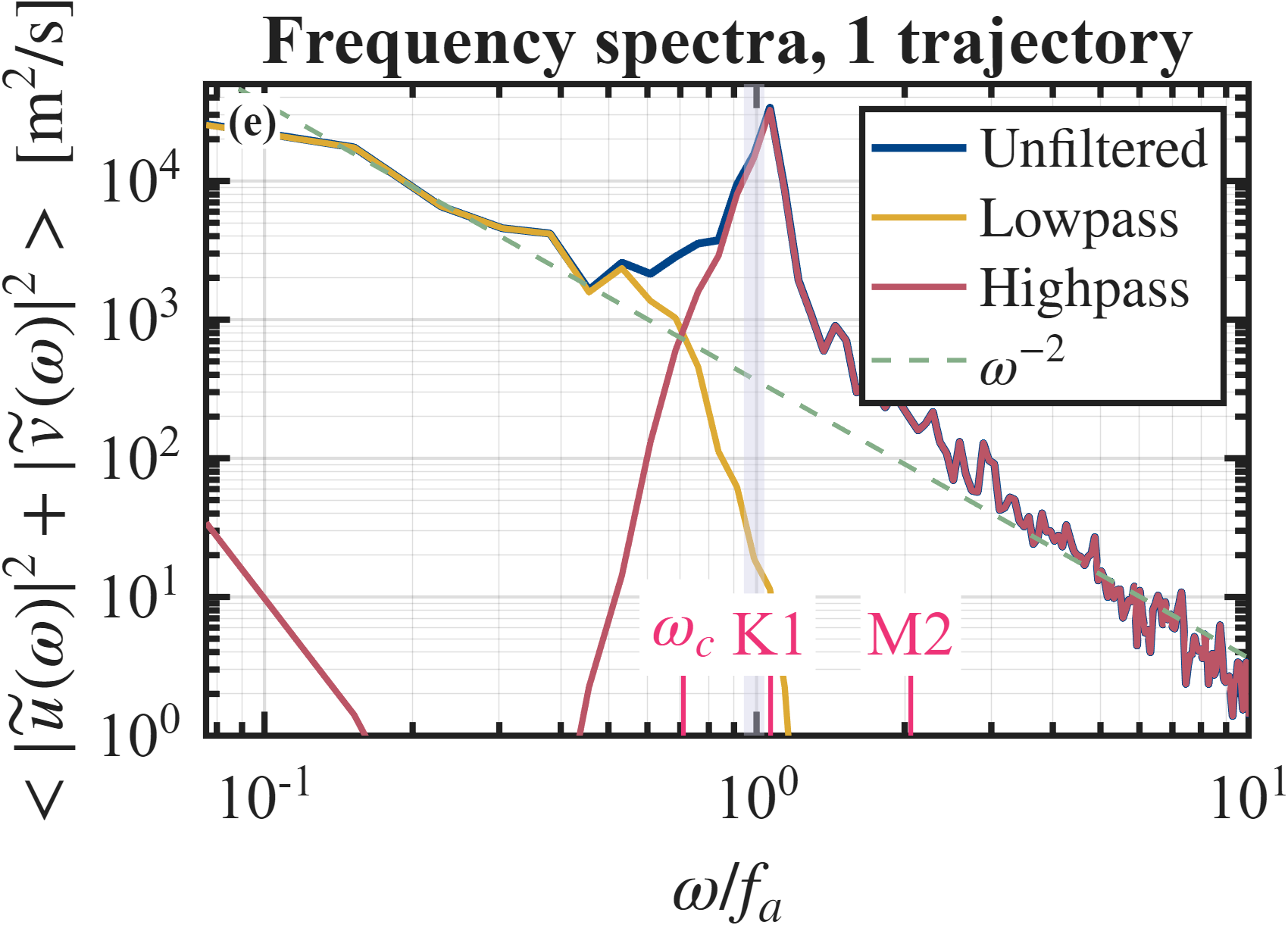}
  \end{minipage}
  \caption{Example with one drifter's 45-day recording from GLAD. Panel (a): position of the sample drifter; line denotes the sample drifter's trajectory and  contour shows observation density of screened data in GLAD. Panel (b):  particle trajectory (``Unfiltered''), which is the sequence of drifter positions from \citep{GLAD15min} directly, and  lowpass trajectory (``Lowpass''), which is the sequence of lowpass filtered positions; the triangle sign marks the drifter deployment position. Panel (c): meridional displacement (difference in meridional position from the deployment position) along the particle and lowpass trajectories. Panel (d): meridional velocity before and after filtering.  Panel (e): frequency spectra before and after filtering; plot labels are the same as Fig.~\ref{fig:GLADLASERoverview}. }
  \label{fig:filtertrajsample}
\end{figure}

We use the processed data products \citep{GLAD15min,LASER15min}, described in detail in \citet{berta2020submesoscale}; aspects relevant to this work are summarized here.
In these data products, 
drifter positions are first interpolated to 5-minute intervals and velocities are estimated by finite differencing. Positions (and hence velocities) are lowpass filtered with a 1-hour cutoff period and resampled to uniform 15-minute intervals. Therefore, motions at periods shorter than 1 hour are absent in our analysis. The data clearly retain imprints from internal waves. In individual drifter trajectories, inertial oscillations are often visible [Fig.~\ref{fig:filtertrajsample}, ``Unfiltered trajectory" in panel (b)]. 
The GLAD product \citep{GLAD15min} contains only CODE-type drifter measurements \citep{davis1985drifter}, while the LASER product \citep{LASER15min} contains only CARTHE-type drifter measurements \citep{novelli2017biodegradable}. 
Both designs are intended to measure near-surface currents; CODE and CARTHE drifters follow the flows in roughly the upper 1~m and 0.6~m of the water column, respectively \citep{davis1985drifter,novelli2017biodegradable}, and have broadly similar responses to wind and surface waves \citep{poulain2022comparing}.

We discard observations flagged as undrogued.
Following \citet{balwada2022direct}, we focus on a region with relatively homogeneous statistics by retaining only offshore observations with longitudes between $91^\circ$W and $84^\circ$W, latitudes north of $24^\circ$N, and seafloor depths greater than 500 m. We also discard snapshots with one drifter observation. After this screening, 1,264,764 and 1,411,873 space--time observations are retained for GLAD and LASER, respectively.
The sampling distribution of screened observations in GLAD is shown in Fig.~\ref{fig:filtertrajsample}(a).

\subsection{Unfiltered SF2s} \label{sec:unfilteredSF2}
The main statistical quantities analyzed in this work are SF2s computed before and after Lagrangian filtering. 
This subsection reviews the unfiltered SF2s and their Helmholtz decomposition. The underlying concepts follow previous works, but we adopt notations that  distinguish unfiltered quantities from the filtered counterparts introduced in \S~\ref{sec:filtSF2}.

At each time snapshot, we identify all the available pairs of drifters. We treat all vectors in a local two-dimensional Cartesian frame. For a given pair, the unfiltered separation vector 
$\hwvec{\ro}$
is the difference between the two unfiltered positions $\hwvec{\xo}^1$ and $\hwvec{\xo}^2$;  the subscript ``$\mathrm{o}$''  denotes ``original'', unfiltered quantities.  
The left panel of Fig.~\ref{fig:sepvector} illustrates $\hwvec{\ro}$ for one drifter pair.

The unfiltered velocity difference between the two drifters, $\hwvec{\duo}(\hwvec{\ro}, \hwvec{\xo}^1,t)$ (only one position is sufficient to define the velocity difference when combined with $\hwvec{\ro}$),
is projected into a longitudinal component $\dulo(\hwvec{\ro},\hwvec{\xo}^1,t)$  along $\hwvec{\ro}$, and a transverse component $\duto(\hwvec{\ro},\hwvec{\xo}^1,t)$  perpendicular to $\hwvec{\ro}$; the subscripts ``L'' and ``T'' stand for ``longitudinal'' and ``transverse'':
\beq \label{dult_a}
\dulo(\hwvec{\ro},\hwvec{\xo}^1,t)\hwdef \hwvec{\duo}(\hwvec{\ro},\hwvec{\xo}^1,t)\cdot\ero,\quad \duto(\hwvec{\ro},\hwvec{\xo}^1,t)\hwdef \hwvec{\duo}(\hwvec{\ro},\hwvec{\xo}^1,t)\cdot \LL(\ez\times \ero\RR),
\eeq
where $\ez$ is the upward vertical unit vector, and $\ero$ is the unit vector along $\hwvec{\ro}$. 

So far we have defined important subscripts ``o'', ``L'', and ``T''; more will appear, and we list in Appendix A key subscripts in this work. 

Assuming weak temporal stationarity and horizontal spatial homogeneity (``stationarity'' and ``homogeneity''), the ensemble-averaged quantities (SF2s and spatial covariances) are functions of separation only, and not functions of the spatial position of one drifter $\hwvec{\xo}^1$ or time ($t$).
The unfiltered longitudinal and transverse SF2s are then defined as
\beq   
\DLO(\hwvec{\ro})\hwdef\avesq{\dulo(\hwvec{\ro}, \hwvec{\xo}^1, t)}, \quad \text{and}\quad \DTO(\hwvec{\ro})\hwdef\avesq{\duto(\hwvec{\ro}, \hwvec{\xo}^1, t)},
\eeq
where $\ave{\cdots}$ denotes ensemble averaging, which is approximated by averaging over all snapshots ($t$), all positions $\hwvec{\xo}^1$, and all drifter pairs whose separation vectors are $\hwvec{\ro}$.
In practice, one counts drifter pairs with separations falling into relevant discrete bins. 

Due to the spatial sparsity of drifter observations, it's impractical to directly evaluate SF2s as functions of two-dimensional $\hwvec{\ro}$. Commonly, isotropy is assumed, so that SF2s are functions of $|\hwvec{\ro}|$ only. In this work, we relax the assumption of isotropy, following \cite{wang2021anisotropic} and expand each SF2 into azimuthal Fourier modes. For example, writing $\ro=(\ro,\alpha_o)$,
\beq \label{Fouriermodes}
\DLO(\hwvec{\ro})=\sum_{n=0}^{\infty} \DXXxn{LLo}{cn}(\ro)\cosb{n\alphao}+\DXXxn{LLo}{sn}(\ro)\sinb{n\alphao}.
\eeq
In the main text, we only analyze the modes at $n=0$, which are angularly independent and related to the two-dimensional functions via an unweighted angle averaging. The leading angularly dependent modes are discussed in Supplementary Materials \S~1. For brevity, outside  Supplementary Materials \S~1, we omit the superscripts $``c0''$ and write, for example, $\DXXxn{LLo}{c0}(\ro)$ as $\DLO(\ro)$.
The separation distance $\ro$ is binned into 26 logarithmically spaced bins from $25$~m to $1052.1$~km. 
Within each bin of $\ro$,  observations are weighted by the distribution of polar angles $\alphao$, to reduce bias from uneven angular sampling under anisotropic statistics.
The specific algorithm we apply is described in Appendix B of \cite{wang2021anisotropic}.

Aside from the angularly weighted averaging, our evaluation of SF2s at $n=0$ is numerically identical to evaluations of SF2 under the assumption of isotropy in other works \cite{balwada2016scale,balwada2022direct}. We stress that we do not assume isotropy, but by evaluating the modes at $n=0$, we are only diagnosing the angularly averaged components of SF2s in the expansion \eqref{Fouriermodes}. 

We focus on \HW{interpreting} the sum of angularly averaged SF2 components
\beq \label{unfilteredSF2}
\DLO(\ro)+\DTO(\ro).
\eeq
We offer an idiosyncratic recount on its connections to flow energetics. We first inspect the two-dimensional SF2s. Manipulating expressions as in Eqn. (18)-(22) of \cite{wang2021anisotropic}, 
\begin{align} 
\DLO(\hwvec{\ro})+\DTO(\hwvec{\ro}) 
&= 2\left[\ave{u^2+v^2}-C_u(\hwvec{\ro})-C_v(\hwvec{\ro})\right]
\\
&=\frac{2}{(2\pi)^2}\left[\ave{u^2+v^2}-\int_{-\infty}^{\infty}[\widehat{C}_u(\hwvec{k})+\widehat{C}_v(\hwvec{k}]\ee^{\ii \hwvec{k}\cdot\hwvec{\ro}}\,dk\,dl\right],\label{twoD_SFsum}
\end{align}
where $(u,v)$ are zonal and meridional velocities, $C_u(\hwvec{\ro})$, $C_v(\hwvec{\ro})$ are covariance functions \HW{defined under homogeneity and stationarity}; e.g.,
$
C_u(\hwvec{\ro}) \hwdef \ave{u(\hwvec{x_1},t)u(\hwvec{\ro}+\hwvec{x_1},t)},
$
and $\widehat{(\cdot)}$ denotes Fourier coefficients in wavenumber space. 
The covariance functions are inverse Fourier transforms of wavenumber power spectra \citep{yaglom1952introduction,buhler2014wave}, so $\widehat{C}_u, \widehat{C}_v$ are power spectra, and their sum is twice the kinetic energy spectrum. Therefore, after averaging \eqref{twoD_SFsum} over the angle $\alpha$, the angularly averaged SF2 sum is related to the angularly averaged horizontal kinetic energy spectral density $E(\kappa)$ [such that $\int_0^{\infty} E(\kappa) d \kappa = \ave{u^2+v^2}/2$ and $\kappa = \sqrt{k^2 + l^2}$ is the radial horizontal wavenumber] via:
\citep{bennett1984relative, davidson2015turbulence, balwada2016scale} 
\beq \label{SF2sum_angularave}
\DLO(\ro)+\DTO(\ro) = 4 \int_0^{\infty} E(\kappa)[1-J_0(\kappa \ro)] d \kappa \approx r_{\mathrm{o}}^2\int_0^{2/\ro} \kappa^2E(\kappa)d\kappa + 4\int_{2/\ro}^{\infty} E(\kappa)d\kappa,           \eeq
where $J_0(\kappa\ro)$ is the zeroth-order Bessel function and the approximation arises from asymptotic behaviors of $J_0(\kappa \ro)$. Thus, the angularly averaged SF2 \eqref{SF2sum_angularave} is contributed by large-scale enstrophy and small-scale kinetic energy, with the weighting controlled by  $\ro$. The algebra in the derivations here are the same as previous works \citep{bennett1984relative, lacasce2016estimating, balwada2016scale} with a slight difference in interpretations: we do not assume isotropy, but rather, we are inspecting the angularly averaged quantities.

If $E(\kappa)$ follows a power law  $\kappa^{-m}$ over an infinite range with $1<m<3$, the corresponding SF2 scales as $\ro^{m-1}$. 
However, if the power law holds over finite ranges, as with real-world data, the correspondence need not be one-to-one \citep{lacasce2016estimating}. So some caution should be exercised if one attempts to use the SF2 to infer the energy spectral power law. 
\HW{Recent work has used additional drifter statistics to validate the spectral-slope inference  \citep{qian2025inferring}. 
In SF2 figures, we show reference slopes $2/3, 1, 2$, but generally refrain from inferring precise spectral slopes from SF2 slopes. }
\subsection{Recipe for Helmholtz decomposition}
\HW{As discussed in the Introduction, the Helmholtz decomposition is widely used to infer dynamical information from velocity observations. Here we list the framework and recipe for the Helmholtz decomposition of SF2s.}
Under appropriate boundary conditions, a two-dimensional velocity field $(u,v)$ can be decomposed into rotational and divergent components induced by a stream function $\psi$ and a potential function $\phi$:
\beq \label{Helmdef}
u(x,y)=\frac{\partial{\phi(x,y)}}{\partial x}-\frac{\partial{\psi(x,y)}}{\partial y}, \quad v(x,y)=\frac{\partial{\phi(x,y)}}{\partial y}+\frac{\partial{\psi(x,y)}}{\partial x}.
\eeq

Although spatially sparse drifter observations do not allow this decomposition to be applied directly to instantaneous velocity fields, the rotational and divergent contributions to SF2s can be diagnosed.
We apply the Helmholtz decomposition formulae for the angularly averaged SF2s \eqref{SF2sum_angularave} :
\begin{align} 
\DRi(\rj) &=\DTi(\rj)+\int_0^{\rj}\LL(-\DLi(\tau)+\DTi(\tau)\RR)/\tau \diff \tau, \label{Helmiso1}\\
\DDi(\rj) &=\DLi(\rj)-\int_0^{\rj}\LL(-\DLi(\tau)+\DTi(\tau)\RR)/\tau \diff \tau, \label{Helmiso2}
\end{align}
where $\DRi(\rj)$ and $\DDi(\rj)$ denote the contributions of the rotational and divergent motions, respectively, to the angularly averaged SF2. These components satisfy 
$\DRO(\ro)+\DDO(\ro)=\DLO(\ro)+\DTO(\ro)$, with no additional cross-term contribution from correlations between rotational and divergent velocities \citep{lindborg2015helmholtz}.
In the subscripts, ``$\psi$'' and ``$\phi$'' denote quantities induced by the rotational and divergent motions respectively. The indices ``i'' and ``j'' are dummy variables; ``i'' can be ``o'' (original), ``f'' (fast), or ``s'' (slow), and ``j'' can be ``o'' or ``s''. The ``f'' and ``s'' subscripts are introduced with the filtered SF2s in \S~\ref{sec:filtSF2}. When evaluating the integrals in \eqref{Helmiso1}--\eqref{Helmiso2}, we use the smallest bin center of $\rj$ as the lower limit of integration.

We refer to \citet{buhler2014wave} and \citet{lindborg2015helmholtz} for the derivations of the Helmholtz decomposition formulae \eqref{Helmiso1}-\eqref{Helmiso2} under isotropy, and to \citet{wang2021anisotropic} and \citet{lindborg2025complete} for the derivations targeting angularly averaged SF2s without assuming isotropy. 
In addition to stationarity and homogeneity, the Helmholtz decomposition of the angularly-averaged modes \eqref{Helmiso1}-\eqref{Helmiso2} requires no further assumptions.
An additional assumption imposed by \citet{wang2021anisotropic} on the cross-correlation between $\psi$ and $\phi$ was later shown to be unnecessary by \citet{lindborg2025complete}.
Supplementary Materials \S~1 discusses the Helmholtz decomposition of anisotropic modes; from GLAD and LASER data considered in this work, we find anisotropic modes inconsequential to the qualitative interpretations of the Helmholtz decomposition.

\subsection{Reference frame for Lagrangian filtering} \label{sec:Refframeexplain}
We approximate drifters as passive tracers following the flow in the top meter of the ocean. This is partially justified as we only use data flagged as drogued. As the drifter observations are horizontal-flow-following, filtering time series along each drifter is equivalent to filtering in the Lagrangian frame of the horizontal flow. 

Our choice of the reference frame is inspired by the generalized Lagrangian-mean (GLM) framework \citep{andrews1978exact,buhler2014b}. 
Many averaging operators can be defined in a Lagrangian frame to compute mean flow \citep{gilbert2018geometric};
here, we define averaging as applying a lowpass filter in time at a fixed drifter label, and regard the highpass and lowpass motions as ``waves'' and ``mean flows'' respectively. This aligns with the GLM requirement that the wave-induced displacement has zero temporal mean to leading order \citep{andrews1978exact,buhler2014b}.  
This choice is  motivated by 
the time-scale separation between balanced and unbalanced motions  in oceanic flows, as discussed in the Introduction. We therefore interpret waves as unbalanced motions and mean flows as balanced motions.
Similar Lagrangian frequency-filtering approaches have been used to decompose waves and mean flows and to diagnose unbalanced and balanced motions in oceanic modeling data, with \citep{kafiabad2022grid,baker_2024_14237745} or without \citep{shakespeare2021new,jones2023using} the GLM framework in mind.

The GLM framework uses the \emph{mean trajectory} as the reference frame.
The mean trajectory is the trajectory due to the mean flow, and waves induce the particle's displacements away from this mean trajectory (hereafter ``wave displacements'').  Variables (e.g., the velocities) associated with both the waves and mean flows use the mean trajectories as their coordinate.
Fig.10.2 in \cite{buhler2014b} provides a visual illustration. 
In our application, this means attributing \emph{both} highpass and lowpass velocities to the \emph{lowpass-filtered trajectory}. 
Even without the GLM framework in mind, this attribution appears natural: for frequency filtering to separate waves and mean flows, the artifacts introduced in the Eulerian frame, i.e., Doppler shifts, are induced by the advection of wave energy by mean flows. 
Canceling this artifact requires an observer moving with the mean flow, whose trajectory is the time integral of the mean-flow velocity, i.e., the lowpass trajectory.
See \cite{gerkema2013note} for a review. 

In principle, different frameworks where the wave and mean flows are attributed to other reference frames can be constructed. One alternative is to attribute filtered velocities to the unaveraged particle trajectories, as in \cite{shakespeare2021new}. We refer to this as the ``particle trajectory framework''. 
The particle trajectory framework may appear natural too: the particle trajectory is directly observed, whereas the mean trajectory under the GLM framework is a conceptual construction. But this choice can complicate the wave-mean decomposition: in the particle trajectory framework, mean flow velocities are assigned to positions displaced by the waves, so deformation of the reference frame (and hence, motions observed from the reference frame) can introduce apparent flow evolutions taking place at wave time scales into the diagnosed mean flow.
Analytically, useful dynamical properties such as the robustness of large-scale geostrophic balance in the mean flow in the presence of strong waves would hold less exactly under the particle trajectory framework \citep{kafiabad2021wave}. 

In our application to drifter observations, the difference between the two frameworks turns out to be significant; in particular, the mean trajectory frame yields a more physically interpretable Helmholtz decomposition of highpass SF2s. We demonstrate this in \S~\ref{sec:lessons}. 


\subsection{Lagrangian-filtered SF2s} \label{sec:filtSF2}
\begin{figure}
         \centering
                  \includegraphics[width=0.9\textwidth]{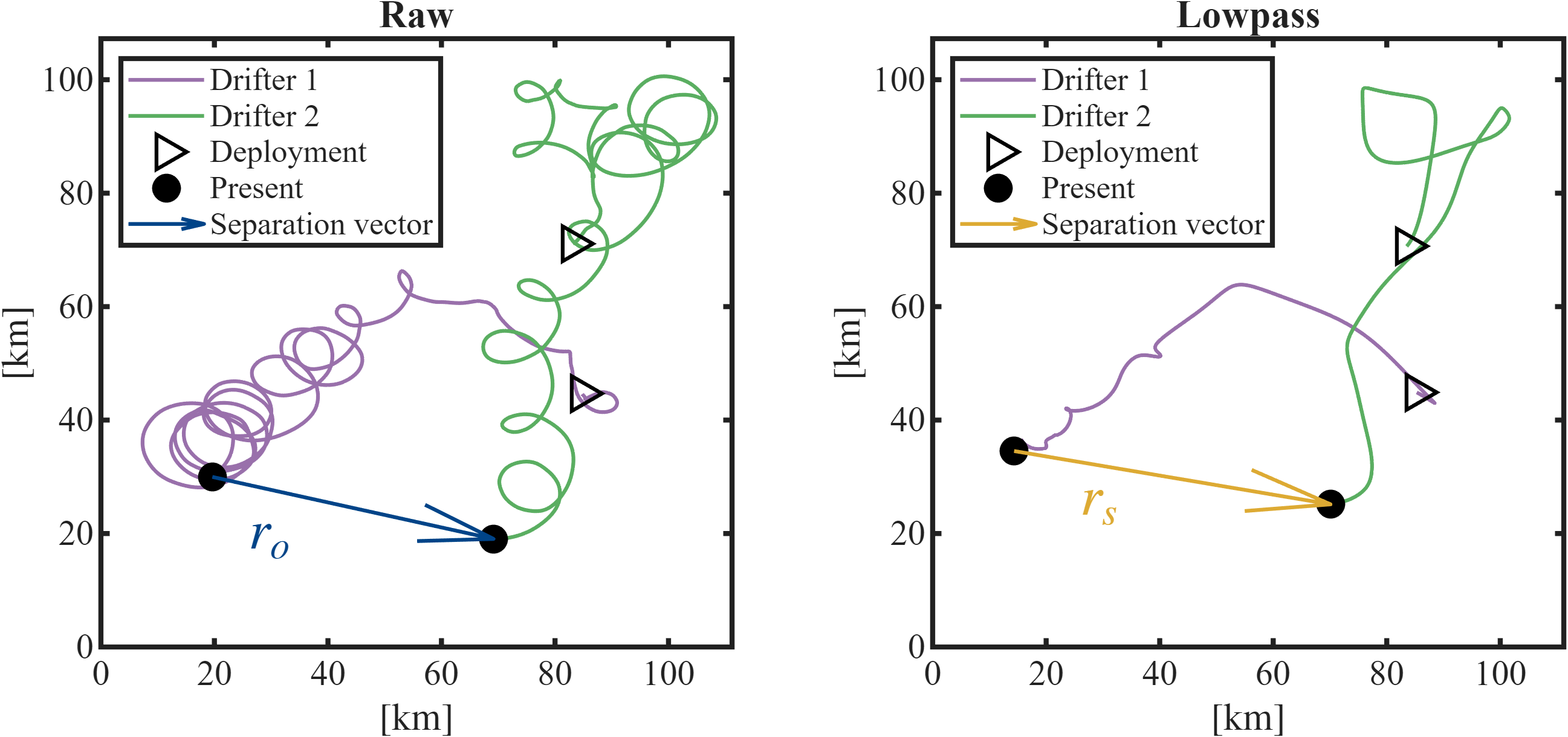}
     \caption{Raw (left panel) and lowpass-filtered (right panel) trajectories (spanning 13.9 days) of two drifters from the GLAD campaign, plotted on local Cartesian coordinates. The triangles denote drifters' deployment positions and the circles denote the raw (left panel) or lowpass-filtered (right panel) positions of the drifters at the present time. The separation vectors $\hwvec{\ro}$ (left panel) or $\hwvec{\rs}$ (right panel) connect the present positions.}\label{fig:sepvector} 
\end{figure}

We now define SF2s computed from Lagrangian-filtered velocities (hereafter ``filtered SF2s''). For a drifter pair, the lowpass separation vector
\beq \label{rs}
\hwvec{\rs}
\eeq
is defined as the vector difference between the lowpass filtered positions of the two drifters ($\hwvec{\rs} = \hwvec{\xs}^1 - \hwvec{\xs}^2$).  The subscript ``s'' denotes ``slow'' (lowpass) quantities. Fig.~\ref{fig:sepvector} (right panel) illustrates $\hwvec{\rs}$. 

The lowpass SF2s are defined analogously to the unfiltered SF2s, but use lowpass velocities and lowpass separation:
\beq \label{lowpassSF2}
\DLS(\hwvec{\rs}) \hwdef \avesq{\duls(\hwvec{\rs},\hwvec{\xs}^1, t)},\quad \DTS(\hwvec{\rs}) \hwdef \avesq{\duts(\hwvec{\rs},\hwvec{\xs}^1, t)},
\eeq
where $\duls$ and $\duts$ are the longitudinal and transverse components of the lowpass velocity difference:
\beq \label{dult_s}
\duls(\hwvec{\rs},\hwvec{\xs}^1, t)\hwdef \hwvec{\dus}(\hwvec{\rs},\hwvec{\xs}^1, t)\cdot\ers,\quad \duts(\hwvec{\rs},\hwvec{\xs}^1, t)\hwdef \hwvec{\dus}(\hwvec{\rs},\hwvec{\xs}^1, t)\cdot \LL(\ez\times\ers\RR),
\eeq
and $\ers$ is the unit vector pointing along $\hwvec{\rs}$.

Highpass SF2s use the highpass velocity differences but are still evaluated as functions of the lowpass separation $\hwvec{\rs}$ (i.e., highpass velocities are attributed to the lowpass trajectories):
\beq \label{highpassSF2}
\DLF(\hwvec{\rs}) \hwdef \avesq{\dulf(\hwvec{\rs},\hwvec{\xs}^1, t)},\quad \DTF(\hwvec{\rs}) \hwdef \avesq{\dutf(\hwvec{\rs},\hwvec{\xs}^1, t)},
\eeq
where $\dulf$ and $\dutf$ denote the longitudinal and transverse highpass velocity differences. The subscript ``f'' denotes ``fast'' (highpass) quantities.
This use of $\hwvec{\rs}$ as the spatial coordinate conforms with using the mean trajectories as the reference frame, as we have argued for in \S~\ref{sec:Refframeexplain}. 

Similar to the treatment of unfiltered SF2s, in the main text, we discuss filtered SF2s only at the angularly-averaged mode $n=0$ in the expansion \eqref{Fouriermodes} and omit the superscripts. For example, $\DTF(\rs)$ refers to $\DXXxn{TTf}{c0}(\rs)$  in the expansion of $\DTF(\hwvec{\rs})$ following \eqref{Fouriermodes}. 
The angular weighting is applied in the computation of filtered SF2s too. The binning of $\rs$ is set identical to the binning of $\ro$. 

The Helmholtz decomposition formulae for the lowpass and highpass SF2s are identical to those for the unfiltered SF2s (\eqref{Helmiso1}-\eqref{Helmiso2}). The Helmholtz decomposition formulae apply unchanged because at no point in the derivations leading to the formulae \citep{buhler2014wave,lindborg2015helmholtz,wang2021anisotropic,lindborg2025complete} would the filtering affect the algebra, which is grounded only in kinematic (and not dynamic) relations.

\subsection{\HW{Choice and effect of frequency filter}} \label{sec:frequencyfilter}
For each drifter record, we apply a 5th-order Butterworth lowpass filter with a 1.5-day cutoff to the zonal and meridional positions and velocities.  
We choose these parameters to strongly suppress motions at frequencies at and above the local Coriolis frequencies covered by the screened data.
The lowpass trajectory is the sequence of the lowpass-filtered positions.  Because the velocities in \citep{GLAD15min,LASER15min} are obtained by finite differencing  positions, filtering and differencing are linear operations that commute; accordingly, the lowpass trajectory agrees (to numerical accuracy; not shown) with the time integral of the lowpass velocity. \HW{This consistency between lowpass velocity and lowpass trajectory aligns with the GLM framework.}
Highpass velocities are computed as unfiltered velocities minus lowpass velocities. \HW{The high-frequency nature of highpass velocities aligns with the GLM condition that wave displacements have zero temporal mean at leading order.}

Fig.~\ref{fig:filtertrajsample} illustrates the filtering for an example drifter.
The anticyclonic loops contributed by inertial oscillations in the unfiltered trajectory largely disappear in the lowpass trajectory [Fig.~\ref{fig:filtertrajsample}(b)]. 
Fig.~\ref{fig:filtertrajsample}(e) shows the sample drifter's frequency spectrum of horizontal velocity variance (hereafter ``frequency spectrum'') before and after filtering.  The frequency spectra are computed as a statistical average of $\left(|\Tilde{u}(\omega)|^2+|\Tilde{v}(\omega)|^2\right)$, where the tilde denotes the Fourier coefficients in the angular frequency space following the convention
$\tilde u(\omega)=\int_{\mathbb{R}} u(t)e^{-i\omega t}\,dt$.
At super-inertial frequencies, the lowpass spectrum is orders of magnitude smaller than the unfiltered spectrum, and exhibits no peaks around Coriolis frequencies, as intended. 

We apply the same filtering to all the screened GLAD and LASER data (\S~\ref{sec:GLADLASER}). For records with gaps, we apply the filter separately to each continuous segment.
The frequency spectra averaged across all drifters before and after filtering are shown in Fig.~\ref{fig:GLADLASERoverview} (lower row), confirming that the lowpass filtered velocities are predominantly subinertial and the highpass filtered velocities are predominantly super-inertial in the overall statistics.  
To estimate these frequency spectra, we partition trajectories longer than 14~days into non-overlapping 14-day segments (discarding remainders). This yields  742 (GLAD) and 616 (LASER) segments.  Hann windows are applied to each segment, before fast Fourier transforms are computed to obtain $\Tilde{u}(\omega)$ and $\Tilde{v}(\omega)$. 
In the computation of SF2s, time segments of all lengths are included, while the data screening  (\S~\ref{sec:GLADLASER}) remains.
As the frequency spectra use only uninterrupted 14-day segments, 
they are based on $77$\% of the GLAD and $50$\% of the LASER data points used for SF2s.
Applying a Welch-like $50$\% overlap does not affect the frequency spectra perceptibly, while extending segments to 28~days does affect the result significantly (not shown) as many shorter trajectories are discarded.
In this work, on the frequency spectra, we only comment on  qualitative behaviors that are insensitive to the length of time segments.

Our filtering operations are close to what is done after synthetic particles are released in previous particle-based Lagrangian filtering packages such as \cite{shakespeare2021new}. The only (yet conceptually important) difference is that we  attribute highpass and lowpass velocities to lowpass  trajectories, as explained in \S \ref{sec:Refframeexplain}. 
(\citet{shakespeare2021new} evaluates filtered velocities on particle trajectories at temporal midpoints.)

\begin{figure}
         \centering
                   \includegraphics[width=0.9\textwidth]
         {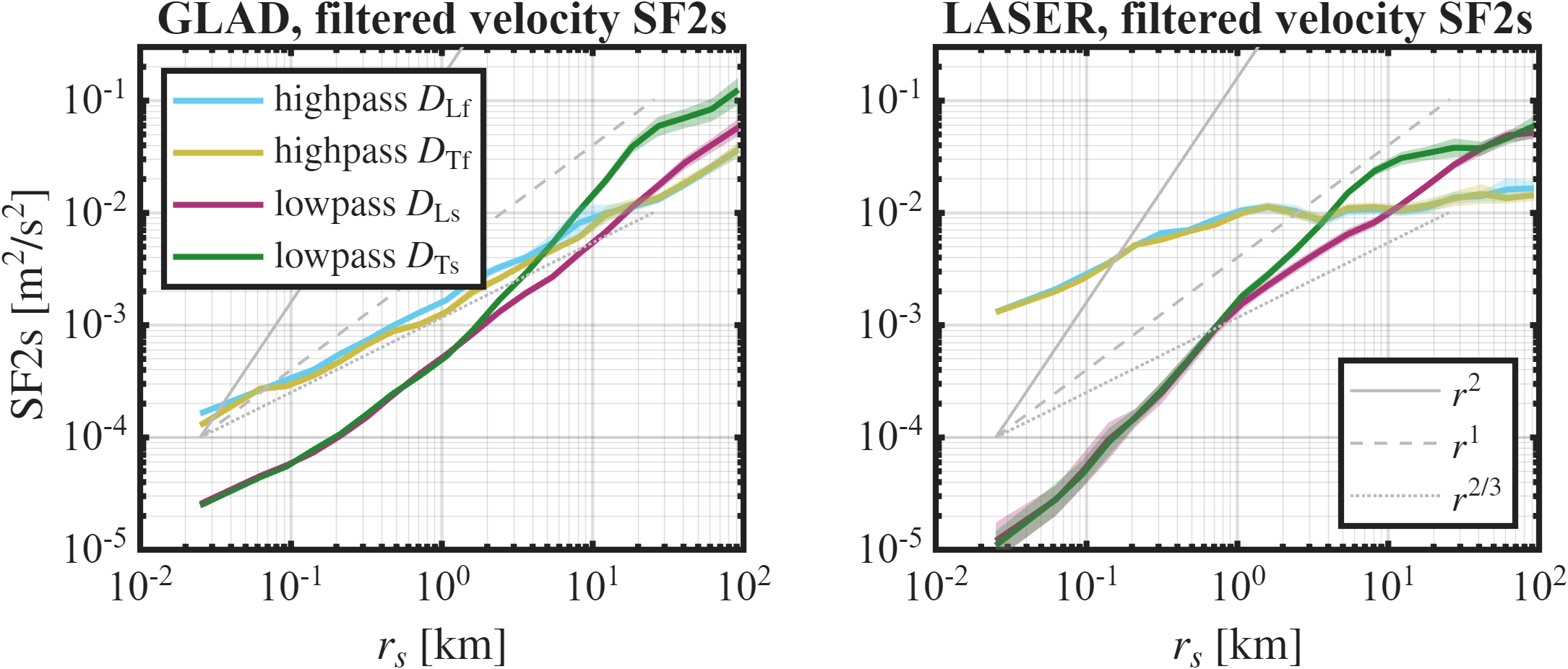}        
     \caption{Filtered longitudinal and transverse SF2s $\DLS(\rs),\DTS(\rs),\DLF(\rs),\DTF(\rs)$ in GLAD (left) and LASER (right). In this and subsequent SF2 figures, shading indicates 95\% confidence intervals estimated via block bootstrapping following \citet{balwada2022direct} (using the same decorrelation time scale). Gray lines mark the power laws $r^2, r^1$ and $r^{2/3}$, with legends in the right panel. 
}\label{fig:dLLdTT} 
\end{figure}

Assembling all the methodological points together, we implement the filtering, compute the filtered SF2s according to \S \ref{sec:filtSF2}, and show in Fig.~\ref{fig:dLLdTT} the angularly averaged, filtered longitudinal and transverse SF2s \eqref{lowpassSF2} and \eqref{highpassSF2}, which are the inputs for the angularly averaged Helmholtz decomposition formulae of filtered SF2s, \eqref{Helmiso1}-\eqref{Helmiso2}. 

\section{Lessons on flow decomposition} \label{sec:lessons}
\subsection{Wave-mean decomposition and pragmatic alternatives} \label{sec:compromises}
\begin{figure}
         \centering
            \includegraphics[width=0.9\textwidth]
         {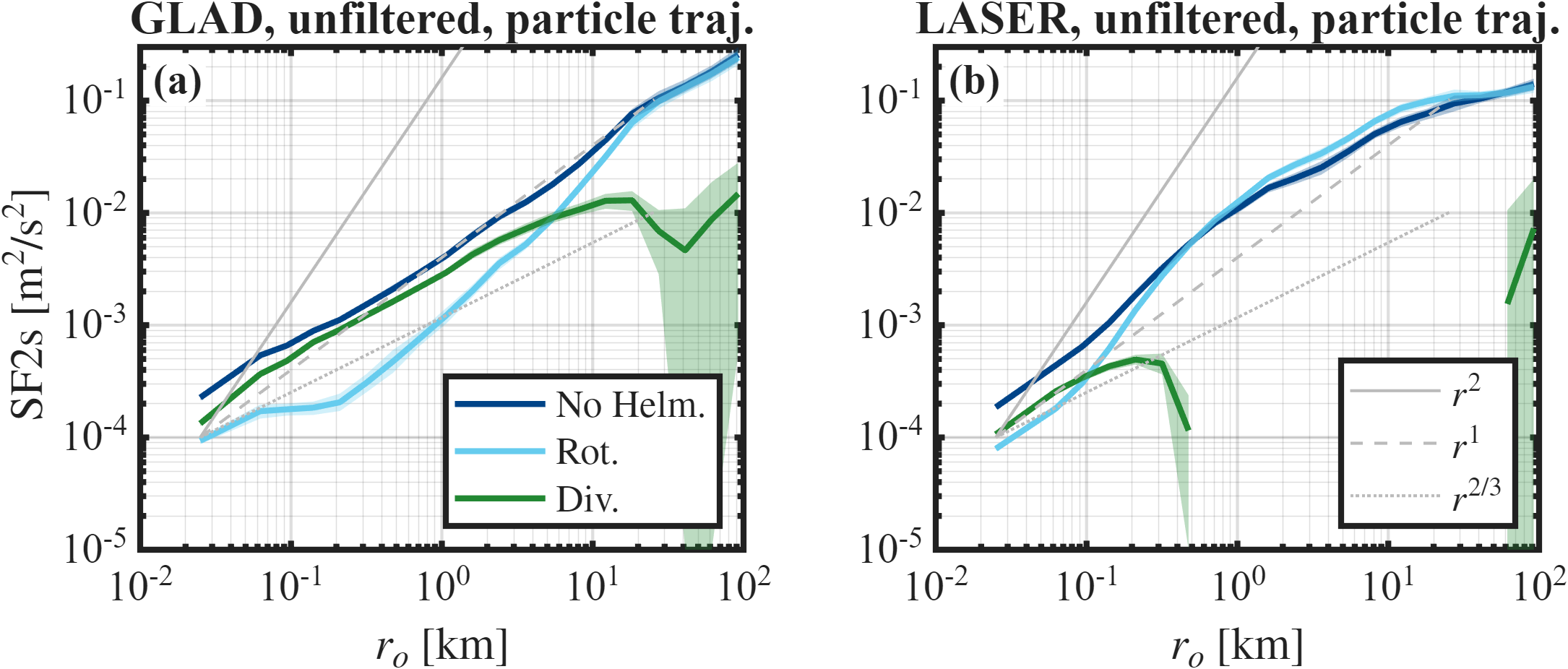} 
   \includegraphics[width=0.9\textwidth]
         {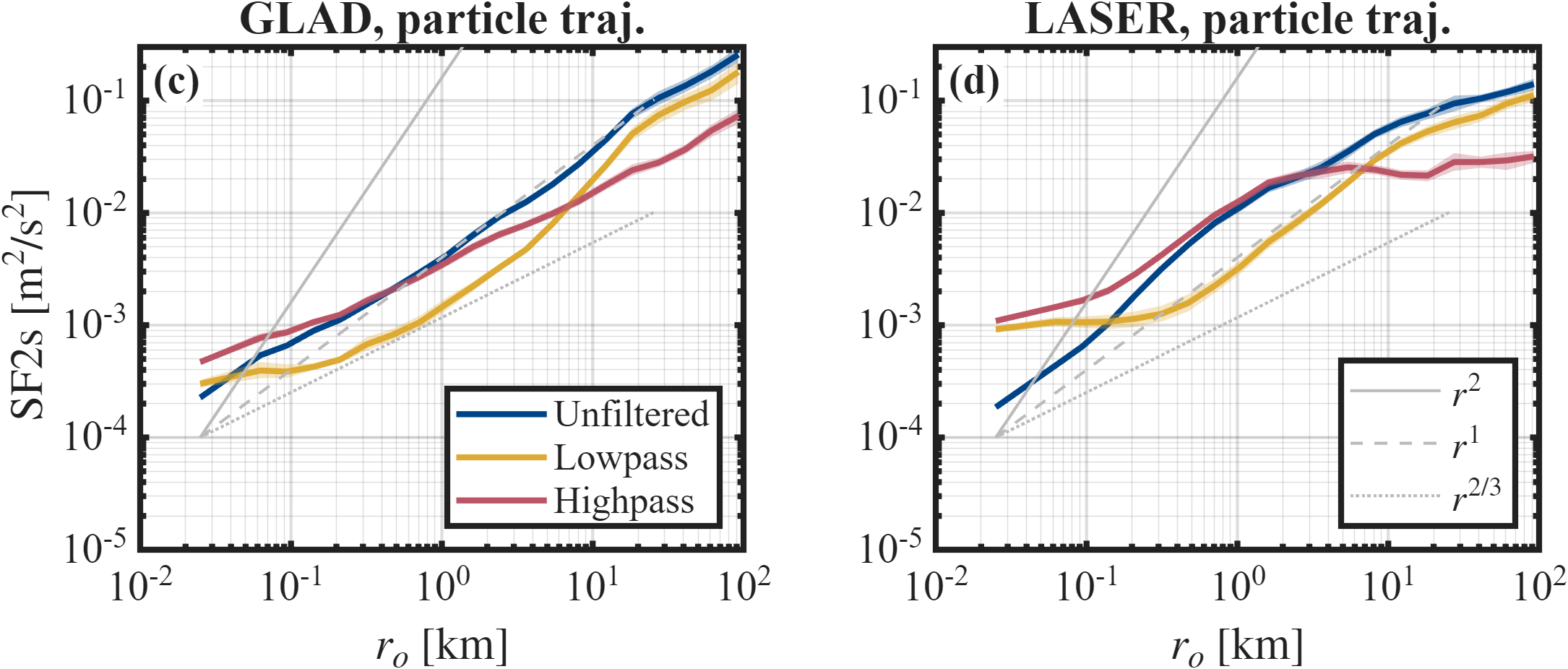} 
   \includegraphics[width=0.9\textwidth]
         {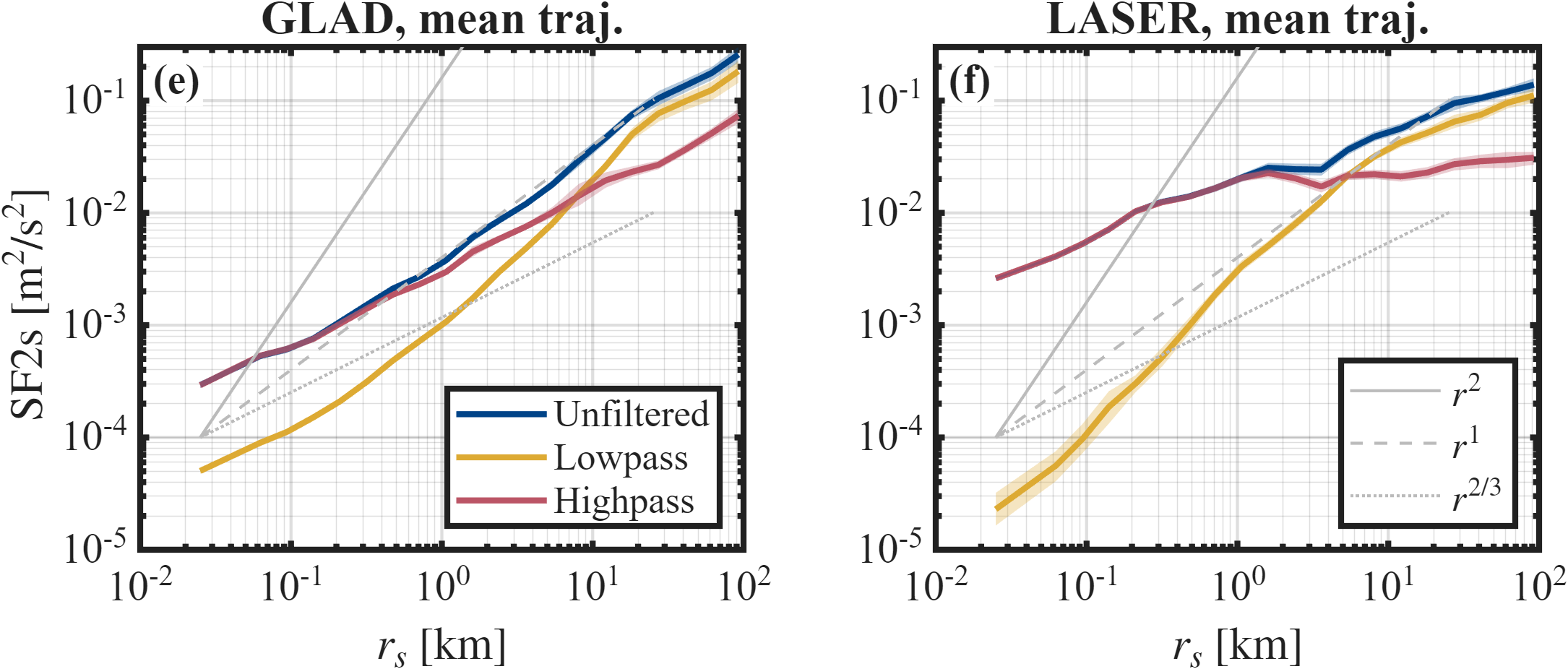}
     \caption{Decomposition of SF2s under different approaches. Lower rows are progressively closer to our preferred wave–mean decomposition.  Left column is from GLAD and right column is from LASER. Panels (a,b): Helmholtz decomposition of unfiltered SF2s , $\left[\DLO(\ro)+\DTO(\ro)\right]$ (``No Helm.'') into rotational $\DRO(\ro)$ (``Rot.'') and divergent $\DDO(\ro)$ (``Div.'') SF2s. Panels (c,d):  
     particle-trajectory-frame filtering, where the ``Unfiltered'', ``Lowpass" and ``highpass" SF2s are all binned over unfiltered separations $\ro$. Panels (e,f): mean-trajectory-frame filtering, where the ``Unfiltered'', ``Lowpass" and ``highpass" SF2s are all binned over mean separations $\rs$. In all SF2 plots, negative SF2 values are skipped; when SF2s are diagnosed positive but the confidence interval covers negative values, shading extends to the bottom of panel.
}\label{fig:compromises} 
\end{figure}

\HWn{Putting together the highpass (wave-induced) and lowpass (mean-flow-induced) SF2s gives, for the first time to our knowledge, a scale-dependent wave-mean decomposition of horizontal kinetic energy under the GLM framework inferred  directly from  observational data [Fig.~\ref{fig:compromises}(e,f)]. As explained in \S \ref{sec:Refframeexplain}, we interpret the highpass and lowpass components as unbalanced and balanced motions, so that this wave-mean decomposition is also interpreted dynamically.}

In Fig.~\ref{fig:compromises}(e,f), the reference ``unfiltered'' curve is the SF2 of the unfiltered velocities binned by the lowpass separation $r_s$:
\begin{equation}\label{unfilteredSF2_on_rs}
\DLO(r_s)+\DTO(r_s).
\end{equation}
Its computation amounts to replacing the lowpass velocities with unfiltered velocities in the computation of the lowpass SF2 \eqref{lowpassSF2}. 
We use \eqref{unfilteredSF2_on_rs} only as a baseline for comparing with the filtered SF2s \eqref{lowpassSF2} and \eqref{highpassSF2}.
When considering unfiltered motion alone, the conventional definition of unfiltered SF2s \eqref{unfilteredSF2} in the particle trajectory frame is more natural.
A more detailed clarification on the usage scenarios between the two definitions of unfiltered SF2s is in Supplementary Materials \S~2. 

From Fig.~\ref{fig:compromises}(e,f), lowpass SF2s are larger than highpass SF2s at $\rs>10$ km in both GLAD and LASER, while at $\rs<5$ km, highpass SF2s are larger than lowpass SF2s. 
Thus, balanced and unbalanced motions dominate large and small separations, respectively, consistent with simplistic textbook pictures (with quasi-geostrophic motions as balanced motions at large scales and internal waves as unbalanced motions at small scales), and with broad patterns found using other decomposition methods in observational and modeling data despite possible deviations from textbook pictures \citep{buhler2014wave,rocha2016mesoscale,qiu2017submesoscale, TorresKlein2018,vladoiu2024energy}. The lowpass SF2s have steeper slopes than highpass SF2s over the diagnosed range.

We next compare this wave-mean decomposition with two pragmatic alternatives, using Fig.~\ref{fig:compromises}(e,f) as the wave--mean decomposition benchmark. First, 
as explained in the Introduction, the Helmholtz decomposition is often used as a heuristic on the wave-mean decomposition. Fig.~\ref{fig:compromises}(a,b) shows the Helmholtz decomposition \eqref{Helmiso1}-\eqref{Helmiso2} of unfiltered SF2s in the particle trajectory frame.
Fig.~\ref{fig:compromises}(a) is quantitatively close to and qualitatively indistinguishable from Fig.~2(a) of \cite{balwada2016scale}, with small differences arising from different data selection and our angle-weighted averaging. 
In GLAD, the Helmholtz decomposition broadly tracks the wave-mean decomposition: the separation where rotational contributions begin to exceed divergent contributions in Fig.~\ref{fig:compromises}(a) is comparable to the separation where lowpass SF2s begin to exceed highpass SF2s in Fig.~\ref{fig:compromises}(e). In LASER, however, the two diagnostics differ qualitatively. 
Fig.~\ref{fig:compromises}(b) suggests strong rotational dominance already for $r_o > 100$~m (with divergent SF2s approaching zero or diagnosed negative due to numerical errors), whereas Fig.~\ref{fig:compromises}(f) shows highpass contributions  comparable to or larger than lowpass contributions over $100$~m--$3$~km, with only moderate  lowpass dominance at larger scales. 
Thus, we argue that Helmholtz decomposition of scale-dependent statistical metrics alone is not sufficient to diagnose the scale-dependent partition between unbalanced and balanced motions; it is most useful when combined with additional information (see Introduction for choices in observational data; in modeling data,  \citet{wang2023simple} used deterministic, spatially resolved information to exploit Helmholtz decomposition to achieve an efficient dynamical decomposition).

Second, we consider a wave-mean decomposition obtained by Lagrangian filtering in the particle trajectory frame [Fig.~\ref{fig:compromises}(c,d)]. \HWn{The particle-trajectory based filtering \citep{shakespeare2021new,jones2023using}, and the filtering of inertial oscillations in \cite{beron2016statistics}, follow this pragmatic alternative.} This alternative is relevant when, for example, a computationally expensive Lagrangian filtering in the particle trajectory frame has already been performed and archived.
In Fig.~\ref{fig:compromises}(c,d), the highpass and lowpass SF2s are computed by replacing the unfiltered velocities by highpass or lowpass velocities in the unfiltered SF2s \eqref{unfilteredSF2}, while retaining the unfiltered separation $r_o$. The unfiltered SF2s in Fig.~\ref{fig:compromises}(c,d) follow \eqref{unfilteredSF2} in the particle trajectory frame.

At separations above $500$~m, in both GLAD and LASER, the particle trajectory frame [Fig.~\ref{fig:compromises}(c,d)] broadly agrees with the mean trajectory frame [Fig.~\ref{fig:compromises}(e,f)] in diagnosing whether the highpass or lowpass SF2s dominate. 
At smaller separations, however, the deviations are large: in the particle trajectory frame [Fig.~\ref{fig:compromises}(c,d)], the diagnosed lowpass or highpass SF2s are often larger than the unfiltered SF2s. This behavior is not observed in the mean trajectory frame [Fig.~\ref{fig:compromises}(e,f)],
where the sum of highpass and lowpass SF2s approximately is equal to  the unfiltered SF2 \eqref{unfilteredSF2_on_rs} (see also Fig.~3 
in Supplementary Materials \S~2).
We explain in Supplementary Materials \S~2 why the approximate summation relationship (highpass SF2 + lowpass SF2 $\approx$ unfiltered SF2) holds better in the mean trajectory frame than in the particle trajectory frame; the main factor is the different extents of phase averaging involved when computing the SF2s in the two reference frames. The more robust summation relationship in the mean trajectory frame is another conceptual convenience (in addition to the cleaner wave-mean separation) over the particle trajectory frame. 
As the differences between mean and particle trajectories arise from wave displacements, which typically happen at small spatial scales, the larger error in the particle trajectory frame in the wave-mean decomposition at smaller separations is expected. 
Therefore,  filtering in the particle trajectory frame can provide a qualitatively correct scale-dependent wave-mean decomposition at spatial scales larger than typical wave displacements. Quantitative error in the particle trajectory frame can nevertheless remain significant even at large scales; for example, in the next subsection, unphysical behaviors of the highpass SF2s in the particle trajectory frame are prominent at all scales diagnosed in LASER.

In summary, using the wave--mean decomposition in the mean trajectory frame as the benchmark, we find that Helmholtz decomposition of unfiltered statistics should not by itself be interpreted as a dynamical decomposition, while filtering in the particle trajectory frame can provide a qualitative indicator of wave--mean partitioning at separations larger than typical wave displacements.

\subsection{Particle trajectory versus mean trajectory} \label{sec:GLMrefframe}
Section \S~\ref{sec:Refframeexplain} explains conceptually why attributing filtered fields to mean trajectories should lead to a cleaner wave--mean separation than attributing them to particle trajectories for Lagrangian filtering.
Synthetic examples have demonstrated this. Fig.~3 of \cite{kafiabad2022grid}, and supplementary movie 1 of \cite{baker2025lagrangian} (comparing panels b and c) are two such examples.  
In these synthetic examples, wave fields are extremely strong, and based solely on these examples, it is still unclear if the choice of reference frame matters in the real ocean. 
Here we show that it does matter, using the Helmholtz decompositions of highpass SF2s in LASER as an example. 
To our knowledge, this is the first observational example in which the mean trajectory frame leads to a more physically interpretable wave--mean separation.

\begin{figure}
\noindent\includegraphics[width=0.9\textwidth]
{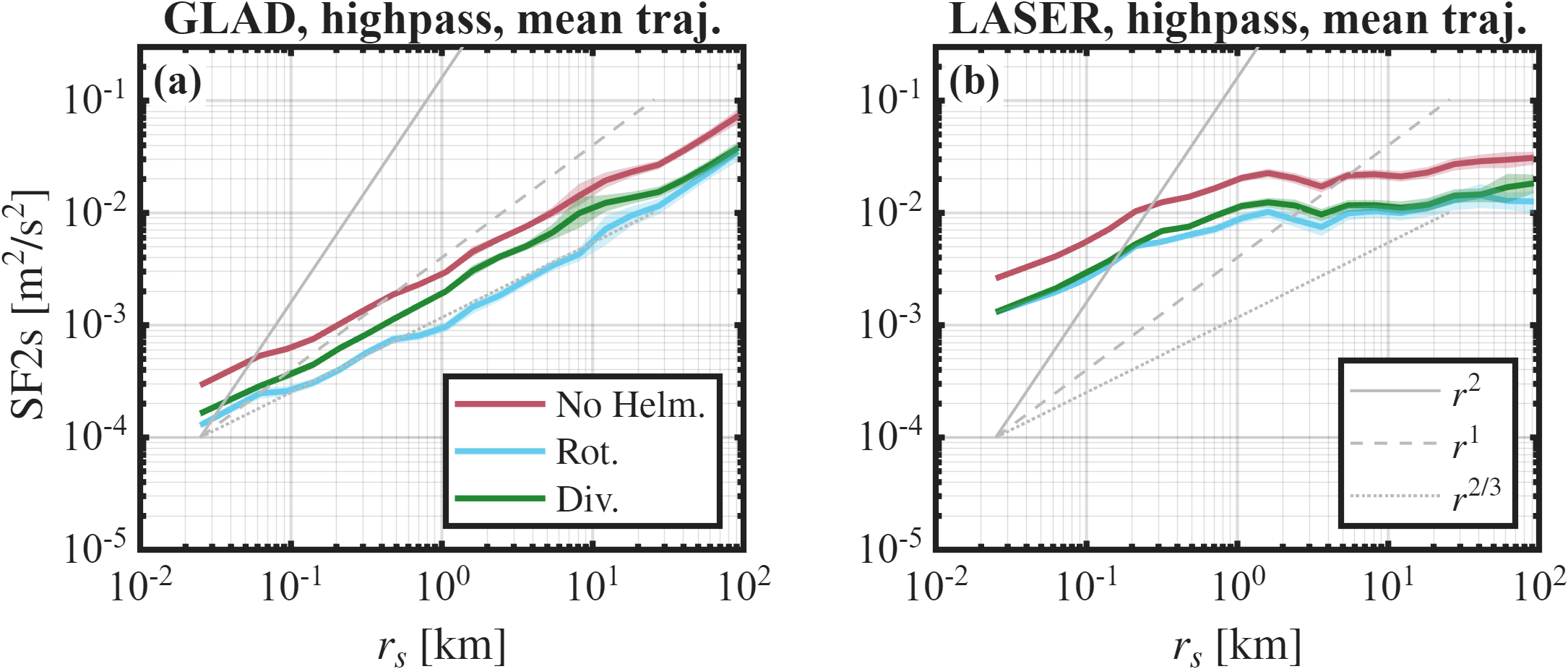}
\noindent\includegraphics[width=0.9\textwidth]{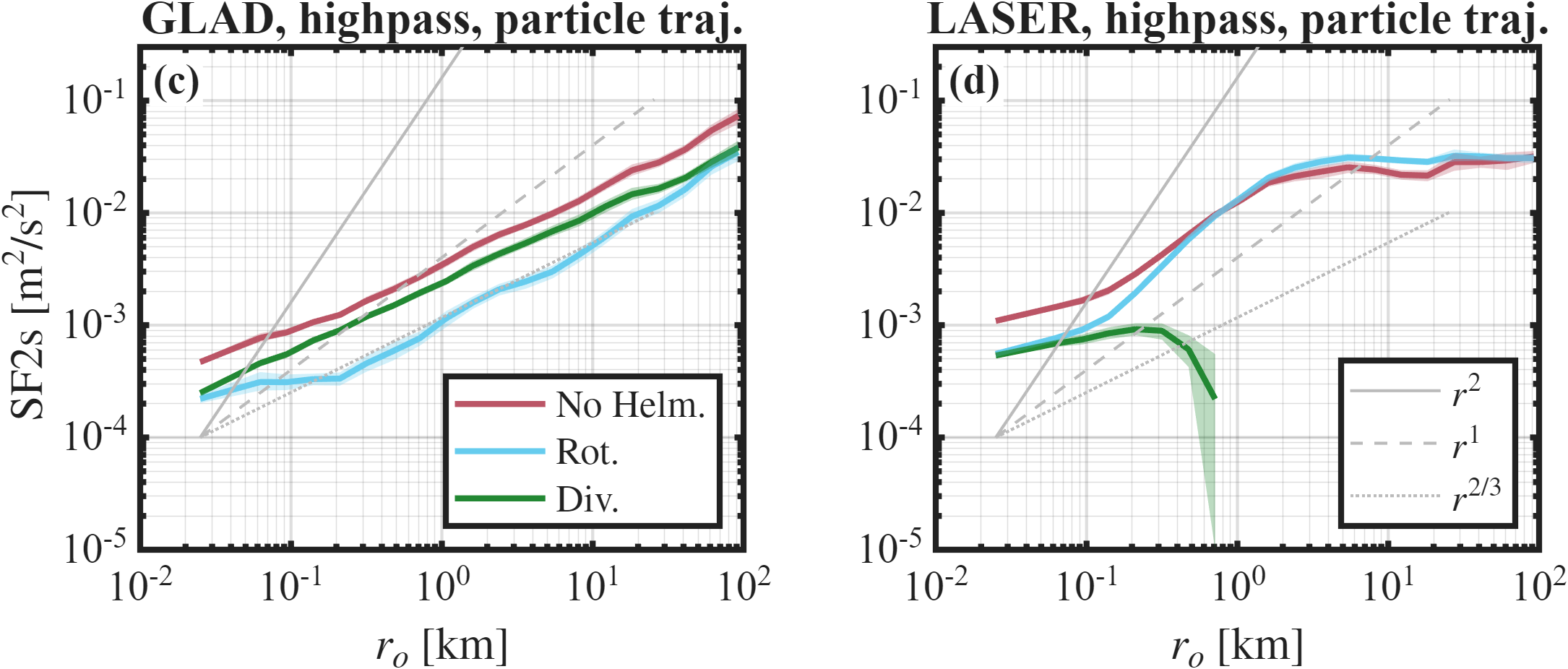}
\caption{Helmholtz decomposition of highpass SF2s computed in two reference frames for GLAD and LASER. Panels (a,c) show GLAD and panels (b,d) show LASER; panels (a,b) use the mean trajectory  frame and panels (c,d) use the particle trajectory frame. We interpret the mean trajectory results (a,b) as providing a cleaner separation of waves from mean flows.}
\label{fig:comparehighpass} 
\end{figure}

Fig.~\ref{fig:comparehighpass} compares highpass SF2s and their Helmholtz decompositions \eqref{Helmiso1}--\eqref{Helmiso2} in the two reference frames. Panels (a,b) use the mean trajectory frame, following our main definition for highpass SF2s \eqref{highpassSF2}. 
Panels (c,d) use the particle trajectory frame; 
there, the highpass SF2s are obtained by replacing the unfiltered velocities by highpass velocities in the unfiltered SF2s \eqref{unfilteredSF2}. 
As the particle trajectory frame does not cleanly remove mean-flow advection effects (\S~\ref{sec:Refframeexplain}), 
we use Fig.~\ref{fig:comparehighpass}(c,d) only as an illustration of possible artifacts.

The contrast in LASER is striking [Fig.~\ref{fig:comparehighpass}(b) versus Fig.~\ref{fig:comparehighpass}(d)]. The two estimates of highpass SF2s differ by about an order of magnitude over part of the separation range, and their Helmholtz decompositions are qualitatively different. In the mean trajectory frame [Fig.~\ref{fig:comparehighpass}(b)], the divergent highpass SF2 is larger than or comparable to the rotational highpass SF2 throughout the diagnosed range, i.e.,
\beq \label{HelmWaveneq}
\DDF(\rs) \geq\DRF(\rs).
\eeq
This inequality is physically explainable. 
Propagating linear oceanic internal waves typically have intrinsic frequencies larger than $|f|$, 
so their divergent kinetic-energy spectrum is at least as large as their rotational kinetic-energy spectrum. See \cite{buhler2014wave}, or Supplementary Materials \S~3 for a recount of the derivation.
The corresponding inequality for angularly averaged SF2s \eqref{HelmWaveneq} follows from the nonnegative kernel $[1-J_0(\kappa\rs)]$ \citep{DLMFch10} 
in \eqref{SF2sum_angularave}.

In contrast, the particle trajectory frame [Fig.~\ref{fig:comparehighpass}(d)] suggests strong dominance of highpass SF2s by rotational component at separations larger than $100$~m, with rotational SF2s exceeding divergent SF2s by more than an order of magnitude.
We are not aware of a model of super-inertial oceanic motions that  predicts such a strong dominance of rotational motions. Rather, this resembles the imprint of nearly geostrophic balanced motions, which are slow and predominantly rotational.
The rotational dominance in Fig.~\ref{fig:comparehighpass}(d) therefore appears to reflect leakage of lowpass motions into the diagnosed highpass SF2s in the particle trajectory frame. This is a manifestation of how the unclean wave-mean separation in the particle trajectory frame can lead to physically uninterpretable diagnostics. 


\section{Dynamical insights in the Gulf of Mexico}
\subsection{Highpass motions}\label{sec:highpass} 
As the highpass motions are dominated by super-inertial frequencies (\S \ref{sec:frequencyfilter}), we interpret the highpass SF2s as imprinted by super-inertial wave dynamics.
Following \S~\ref{sec:filtSF2} and \S~\ref{sec:GLMrefframe}, we only use the mean-trajectory-frame diagnostics in Fig.~\ref{fig:comparehighpass}(a,b), which avoid the mean-flow contamination evident in the particle trajectory frame and provide  better physical interpretability.

The variance of highpass velocities $\ave{\LL(u_f-\ave{u_f}\RR)^2+\LL(v_f-\ave{v_f}\RR)^2}$, computed from all samples used in the highpass SF2 estimates, is significantly larger in GLAD ($0.0682$~m$^2$~s$^{-2}$) than in LASER ($0.0275$~m$^2$~s$^{-2}$). As a reminder, GLAD is in summer and LASER is in winter. 
Plausible contributors include energetic forcing during Hurricane Isaac (included in our processed record), and enhanced surface intensification of internal waves in summer associated with a shallower mixed layer \citep{d1978mixed,rocha2016seasonality,lahaye2019sea}.

In both GLAD and LASER, the highpass SF2s (and the rotational and divergent components) 
show an approximate \(r_s^{2/3}\) power law below their decorrelation scales.
In GLAD, this power law extends over nearly 4 decades of length scales, while in LASER, it spans over about one decade  before the SF2s flatten (indicating decorrelation at \(O(1)\) km).
As there is no strong sign of a different power law regime in the highpass SF2s in the diagnosed ranges, the connection of power laws of these SF2s to power laws of wavenumber spectra is less ambiguous (\S \ref{sec:unfilteredSF2}): the highpass SF2s can be consistent with a $\kappa^{-5/3}$  wavenumber spectrum at spatial scales below the decorrelation scale. Some care is still needed when interpreting this power-law behavior; we expand on this in \S \ref{sec:discussions}. 

\begin{figure}
         \centering
          \includegraphics[width=0.6\textwidth]
         {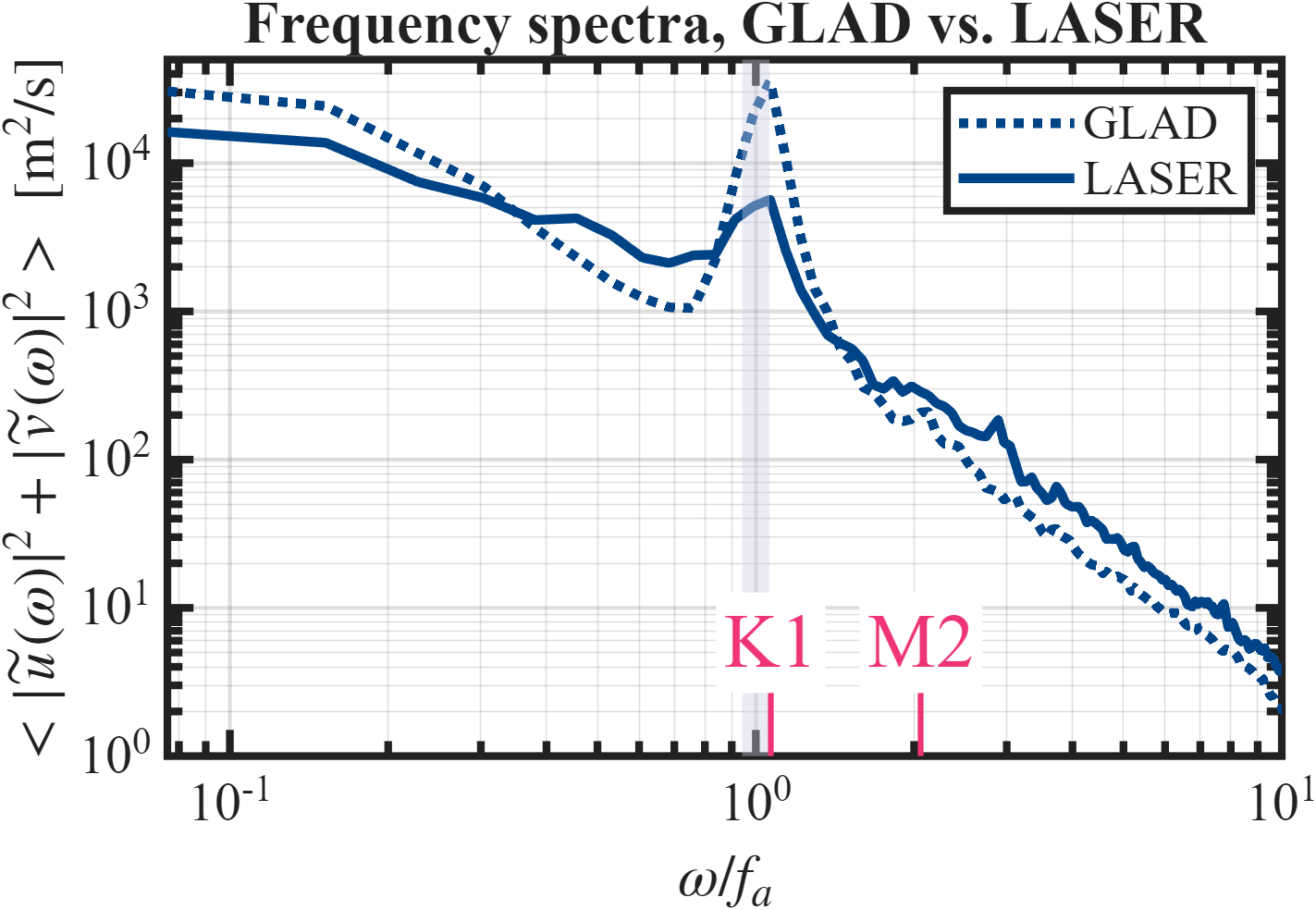}
     \caption{Frequency spectra as in the lower row of Fig.~\ref{fig:GLADLASERoverview}, shown here with GLAD and LASER overlaid for direct comparison.}
\label{fig:freqspec_together} 
\end{figure}

\HWn{Near-inertial peaks are clear in the Lagrangian frequency spectra in LASER and  GLAD (Fig.~\ref{fig:freqspec_together}). 
A bandpass filter over $0.8 f_a -1.2 f_a$ gives velocity variances of $0.0336$~m$^2$~s$^{-2}$ in GLAD and 
$0.0096$~m$^2$~s$^{-2}$ in LASER, corresponding to  
49\% and 35\% of the highpass velocity variance in GLAD and LASER respectively.
This confirms that the near-inertial band is one of the most energetic frequency bands in the highpass field and can plausibly dominate the highpass SF2s over significant ranges of spatial scales. 

The inequality \eqref{HelmWaveneq} observed in the Helmholtz decomposition of highpass SF2s in both GLAD and LASER is consistent with linear internal waves (\S \ref{sec:GLMrefframe}). At large separations ($\rs \gtrsim 10$~km), the Helmholtz decomposed components $\DDF$ and $\DRF$ become comparable  in both LASER and GLAD, consistent with near-inertial waves (``NIWs''), for which $\omega \approx f$.  As  $\omega\approx f$, the divergent and rotational kinetic energy spectra turn equal (Supplementary Materials \S~3), and thus the inequality in \eqref{HelmWaveneq} approaches equality, leading to a near equipartition between rotational and divergent SF2s.  
Because of the dispersion relationship,  the horizontal spatial scales should be much larger than vertical spatial scales in NIWs \citep{gill2016atmosphere}. Therefore, the tendency of $\DDF$ and $\DRF$ to be close at $\rs \gtrsim 10$~km is consistent with a dominance of NIWs at large horizontal scales. 

Indulging further in the NIW-related speculations motivated by the near-equipartition between divergent and rotational highpass SF2s, we note that 
in LASER (winter), the near-equipartition  extends to smaller separations ($\rs \lesssim 10$~km), suggesting that the highpass motions at these scales are also strongly imprinted by NIWs. This is not observed in GLAD (summer). One interpretation is that LASER contains more NIWs at small horizontal scales, which might arise under stronger wave--mean interactions that transfer NIW energy to higher wavenumbers. Consistent with this conjecture are the following observations.
First, the spatial decorrelation scale of highpass SF2s [the separation distance above which SF2s are plateauing in Fig.~\ref{fig:comparehighpass}(a,b)] is $O(1)$ km in LASER, while larger than $O(100)$ km in GLAD (as the plateauing is not present at the diagnosed range). 
This is consistent with wave energy being more concentrated at small scales in LASER.  
Second, 
at submesoscale spatial range ($500$~m -- $10$~km), the lowpass (balanced) motions in LASER are more active than in GLAD (\S \ref{sec:lowpass}); at $\rs\approx1$~km, the lowpass SF2s in LASER are 2-3 times as big as the lowpass SF2s in GLAD (Fig.~\ref{fig:SF2Helmsslow}).
Such balanced motions in LASER can more effectively scatter, refract, strain, or diffuse NIW energy toward smaller scales \citep{bartello1995geostrophic,young1997propagation,buhler2005wave,asselin2020refraction,danioux2016near}.}

This interpretation is supported by circumstantial evidence at best. 
When we use the term ``consistent with'', we imply that the noted observations are only compatible with the suggested theoretical models, and are not direct proofs. 
Diurnal tides are difficult to separate from NIWs as the K1 frequency is close to $f$ in this region, and we do not strictly rule out significant tidal contributions within the inertial band.
Our current emphasis on NIWs is motivated by (i) the weak semi-diurnal tidal peaks in Fig.~\ref{fig:freqspec_together} \HW{(around the marked M2 frequency)}, suggesting tides can be modest overall, 
and (ii) independent observations suggesting that diurnal tides are weaker than NIWs in summer offshore north-east GoM  \citep{gough2016resonant}. 
Other mechanisms can also be consistent. One such possibility is that in LASER, more waves may be generated from fronts through loss of balance \citep{nagai2015spontaneous,molemaker2005baroclinic,shakespeare2014spontaneous}. 
To clarify the dynamical origins, additional data (from independent observations or models) are required.  


\subsection{Lowpass motions} \label{sec:lowpass}
\begin{figure}
         \centering
                   \includegraphics[width=0.9\textwidth]
         {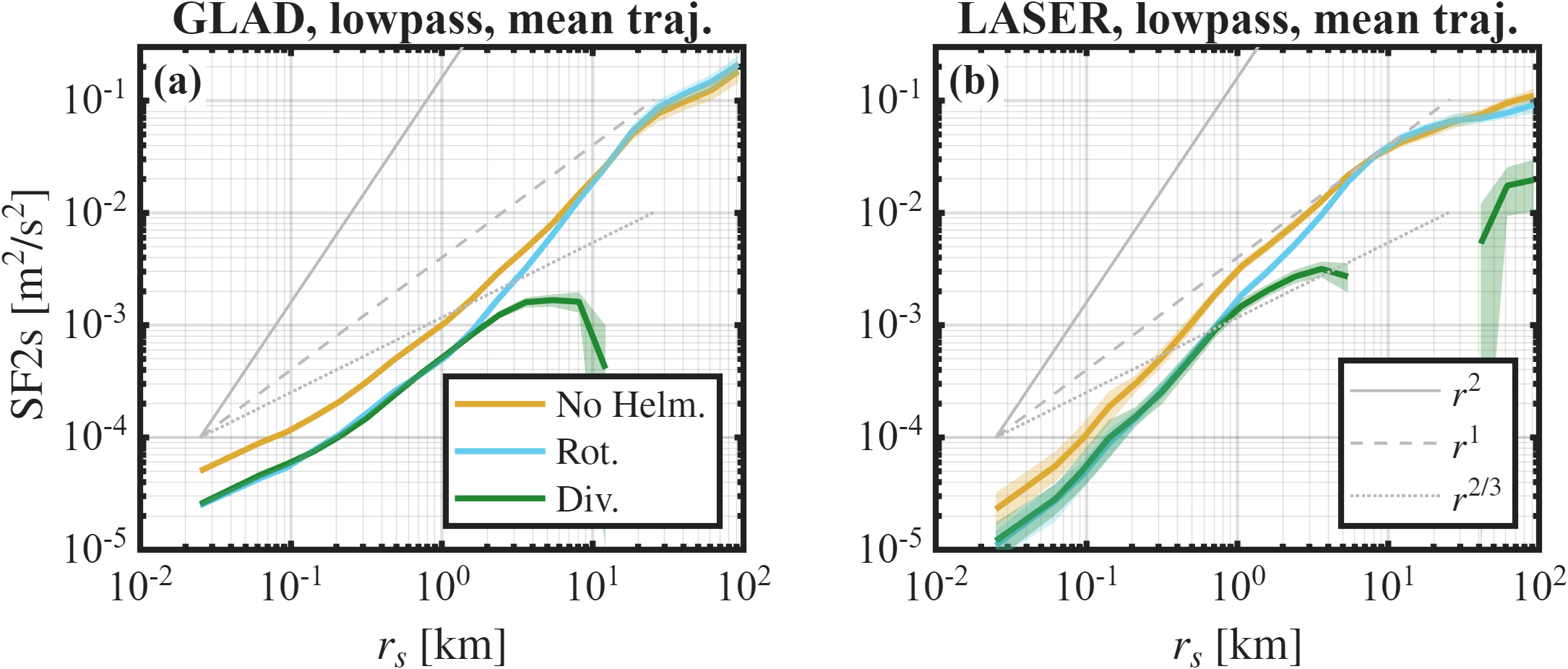}        
     \caption{Helmholtz decomposition of lowpass SF2s computed in the mean trajectory frame from GLAD [panel (a)] and LASER [panel (b)]. 
}\label{fig:SF2Helmsslow} 
\end{figure}

The lowpass SF2s and their Helmholtz decomposition are shown in Fig.~\ref{fig:SF2Helmsslow}. 
In both GLAD and LASER, rotational lowpass SF2s are at least as large as divergent lowpass SF2s across the diagnosed range. At $\rs\gtrsim1$~km, the rotational SF2s are generally at least an order of magnitude larger. The divergent SF2s at these scales are sometimes estimated to be slightly negative as the true values are near zero and sensitive to possible numerical errors in the evaluation of \eqref{Helmiso1}--\eqref{Helmiso2}.
This dominance of rotational lowpass SF2s at large scales 
is consistent with large-scale geostrophically balanced motions and is sharper than the corresponding result for unfiltered SF2s noted in previous works \citep{balwada2016scale}: the dominance of rotational motions is now attributable specifically to the mean flows, whereas the highpass SF2s arising from waves show the opposite tendency or near-equipartition [Fig.~\ref{fig:comparehighpass} (a,b)].

Remarkably, at smaller separations ($\rs<1$~km),  over more than a decade of separation, the lowpass rotational and divergent SF2s are nearly equipartitioned in both GLAD and LASER.

Previous Lagrangian observations have found evidence for strong horizontal divergence and associated vertical motion at submesoscales \citep{dasaro2018ocean,berta2020submesoscale,tarry2021frontal}.
In particular, \citet{dasaro2018ocean} observed persistent tendencies of drifters to converge and pointed out that divergent mean flows are active in LASER; our results here offer a quantification.
In Eulerian observations, \citet{freilich2023characterizing} found strong divergence at scales under $1$~km from airborne snapshots in frontal regions, and \citet{taylor2026seasonal} found an equipartition of the cross-scale kinetic-energy transfer rate due to rotational and divergent motions at $7$~km from a decade of coastal radar observations.
For context, estimates of divergence based on unfiltered velocities can contain large contributions from internal waves, whose motions are dominantly oscillatory, and need not contribute to sustained vertical transport relevant for impacting climatological tracers \citep{balwada2018submesoscale}. The previous observational works discussed here link observed divergence to persistent, mean-flow associated processes through convergence of particles, diagnosed vertical motions, or the dynamical contexts, but they do not quantitatively isolate the mean-flow contributions as we do here. 
We are not aware of an observational precedent for such a scale-extended rotation--divergence equipartition in surface mean flows as we find here.

The kinematic origin of the equipartition is clear: from \eqref{Helmiso1}-\eqref{Helmiso2}, $\DRS\approx\DDS$ when $\DLS\approx\DTS$ over the same and smaller scales, as observed in  Fig.~\ref{fig:dLLdTT}.
The dynamical origin of this equipartition is less clear.
Several processes could plausibly contribute. 
Scaling arguments and modeling results suggest that submesoscale fronts can produce divergent motions that are statistically as active as rotational motions in the balanced flow \citep{barkan2019role}.
Wind-driven Ekman currents may also have strong divergence in sub-inertial motions \citep{gill2016atmosphere}.
However, to our knowledge, existing theories on frontogenesis or Ekman forcing only explain how divergent motions can be strong and locally comparable to rotational motions, but they do not explicitly predict the equipartition across an extended range of spatial scales. 
Another candidate  is the parametric subinertial mixed-layer model of \citet{young1994subinertial}, 
where a rotation--divergence equipartition arises if the inverse vertical-mixing time scale happens to be close to $|f|$; however, this leaves open why the mixing time scale would behave in the speculated way. 
Finally, idealized modeling of stratified turbulence reports approximate rotational--divergent equipartition without prominent waves in the continuous inertial range, even under purely rotational forcing \citep{lindborg2007stratified}; we are not aware of a clear analogy of this model in surface ocean dynamics yet.
To clarify the dynamical origin of this intriguing equipartition, more observational/modeling data and process-oriented studies are required.

We cannot fully exclude that the equipartition might arise from nonphysical artifacts, especially given the small SF2 amplitudes at the small separations. Potential contributors  include (a) spectral leakage from the frequency filter, and (b) the constraint that all SF2s vanish at $\rs=0$ by construction, which could impose  SF2s to be close to each other at small separations.
On (a), as a basic sensitivity test, 
the equipartition persists when we change the Butterworth filter order to be between 1 and 4 (not shown; we do not increase the order over 5 due to apparent numerical instabilities). This suggests some robustness to spectral leakage. 
The artifact (b) should typically only affect the first few separation bins [e.g., Fig.~\ref{fig:compromises}(a), where the rotational and divergent SF2s seem closer in a few bins near 0.01 km than near 0.1 km], and is unlikely to fully explain the equipartition over more than a decade of $\rs$.

The similar partitioning of rotational and divergent SF2s between GLAD and LASER in the lowpass components, despite differences in season, deployment patterns and quantity of available measurements, is remarkable. 
The GLAD and LASER lowpass SF2s still differ in the following  ways. 
LASER has a slightly larger lowpass velocity variance ($0.0116$~m$^2$~s$^{-2}$) than GLAD ($0.0109$~m$^2$~s$^{-2}$), and has lowpass SF2s one to two times larger than GLAD at $500$~m$\leq\rs\leq10$~km.
This tendency of mean flows to be more energetic and more concentrated at submesoscales in winter is consistent with the seasonality of mixed layer energetics under mechanisms such as available-potential-energy-release \citep{boccaletti2007mixed, callies2015seasonality, rocha2016seasonality}. The power-law behaviors are qualitatively different too. 
GLAD transitions between approximate $\rs^{2/3}$, $\rs^2$, and $\rs^1$ regimes,
whereas LASER shows an approximate $\rs^{2}$ power law up to $10$~km before decorrelating.
The absence of a transition in power law regimes in LASER's lowpass SF2s, as compared to GLAD's lowpass SF2s, indicates different scale-dependent mean-flow dynamics between GLAD and LASER.



\section{Summary and Discussion} \label{sec:discussions}
In this work, we apply a GLM-inspired Lagrangian filtering algorithm to surface drifter data to decompose wave and mean-flow contributions to scale-dependent kinetic energy. 
We report the implementation and design considerations of the filtering approach, and analyze its impact on SF2s through Helmholtz decomposition. The main contributions are twofold.
\begin{itemize}
    \item Flow decomposition. To our knowledge, this is the first GLM-inspired wave--mean decomposition applied to oceanic observational data. Physically meaningful results are obtained, and comparisons with particle-trajectory-reference-frame-based decomposition and Helmholtz decomposition are discussed extensively, demonstrating the advantages and applicability of GLM-inspired filtering under real oceanic scenarios.  
    \item Upper-ocean dynamics. We find on the ocean surface 1) signatures of inertial-band waves, which appear active below $1$~km in winter; 2)  divergent motions in the mean flows that are as strong as rotational motions at spatial separations smaller than $1$~km; and 3) a transition from dominance of waves to dominance of mean flows at spatial separations of $5-10$~km in summer and winter. These results inform submesoscale processes such as wave-mean interactions and frontogenesis in the GoM, and motivate further process-oriented studies.    
\end{itemize}

A direct application of our methodology is validation of satellite altimetry products. Lowpass velocities evaluated in the mean trajectory frame can serve as an independent reference for balanced velocities inferred from satellite data, including SWOT sea-surface-height products \citep{tranchant2025swot, bertrand2025robust}. For geostrophic velocity inference, our methodology identifies scales where 
two  necessary conditions for geostrophic dominance are met: 1) lowpass SF2s dominate highpass SF2s, and 2) rotational lowpass SF2s dominate divergent lowpass SF2s.
In GLAD and LASER, both conditions are met at only around $\rs\gtrsim10$~km, suggesting that geostrophic velocities do not dominate at smaller spatial scales in summer/winter offshore GoM.

Our quantification of divergent kinetic energy in the mean flows is also important for interpreting surface divergence as an indicator of vertical transport. 
Large surface divergence in unfiltered motions is not, by itself, evidence of sustained vertical transport: internal waves can produce large instantaneous divergence, but their motions are dominantly oscillatory and may contribute little to irreversible tracer transport \citep{balwada2018submesoscale}. 
The divergence most relevant for sustained vertical exchange of tracers is instead the divergence of the mean flow at low frequencies, such as that associated with convergent secondary circulations near fronts \citep{Mcwilliams2016, barkan2019role,jones2023using,taylor2023submesoscale}. 
This transport-relevant divergence can be much weaker in amplitude than wave divergence and difficult to extract when spatial resolution is coarse or temporal sampling is sparse, which is typical for Eulerian observations. 
Even in fine, time-resolving Eulerian observations, wave--mean separation would rely on Eulerian filtering and therefore remain affected by Doppler shifts \citep{gerkema2013note,callies2020time}.
Lagrangian observations, such as drifters, provide Lagrangian wave-resolving time series spanning a wide range of spatial scales, allowing GLM-inspired Lagrangian filtering combined with Helmholtz decomposition to quantify the scale-dependent divergent kinetic energy contributed by mean flows.

 
The magnitude relationship between rotational and divergent imprints in lowpass SF2s is similar between GLAD and LASER (\S \ref{sec:lowpass}).
In contrast, the Helmholtz decomposition of highpass SF2s differ more substantially between GLAD and LASER  (\S \ref{sec:highpass}).
Thus, the qualitative differences between GLAD and LASER in the unfiltered SF2s and their Helmholtz decompositions noted in previous works \citep{balwada2016scale,balwada2022direct} appear  to be primarily induced by differences in wave statistics.
This also suggests that oceanic states with broadly similar mean-flow statistics can nevertheless have very different wave statistics.

We summarize the assumptions and sources of error of our analysis as follows. 

The GLM-inspired filtering itself is diagnostic, but its dynamical interpretation relies on attributing highpass and lowpass motions as unbalanced and balanced flows, which is based on time scale separation. 
The SF2 estimates use horizontal homogeneity and temporal stationarity when replacing ensemble averages by spatial and temporal averages. We do not assume horizontal isotropy, and focus on the angularly-averaged modes in the main text.  The Helmholtz decomposition of angularly averaged SF2s requires no further assumptions. 

Our analysis relies on the processed data products \citep{GLAD15min,LASER15min}, and  treats drogued drifters as passive tracers of near-surface currents. Measurement error, effects of surface waves and wind drag, and imperfect flagging of drogue loss are ignored. Some of these simplifications are partially justified:
\cite{novelli2017biodegradable} reported that  $95\%$ of the LASER positions have an error of $10 \textrm{ m}$ or less, and drogue loss has been carefully detected by transmission data and comparison of neighboring drifter velocities \citep{haza2018drogue}.

One more systematic source of error is spectral leakage from numerical frequency filters. 
Supplementary Materials \S~2 shows that the difference between unfiltered SF2s and the sum of highpass and lowpass SF2s, which is contributed partly by spectral leakage, is small relative to the unfiltered SF2s.  However,  leakage errors may still be significant relative to the filtered SF2s, which are sometimes much smaller than the unfiltered SF2s. 
We have not devised a way to separately estimate leakage of super-inertial motions into lowpass SF2s and subinertial motions into highpass SF2s.

Statistical uncertainty is represented by 95\% bootstrapped confidence intervals computed with modified block bootstrapping following \cite{balwada2022direct}, which assumes independence between observations separated by a typical flow time scale (taken the same as in \cite{balwada2022direct}). This assumption is still restrictive: two observations in a same convergent zone can be separated by a long time interval and still be statistically dependent.
A fundamental improvement of statistical error quantification is beyond the scope of this work. 

Recent work by \cite{hypolite2026second} provides complementary diagnostics. Using simulations, they apply Eulerian frequency filtering to surface velocities, 
and interpret the resulting highpass and lowpass SF2s partly through Helmholtz decomposition. 
Their super-inertial motions tend to drive the transverse-to-longitudinal SF2 ratio toward unity at submesoscales, consistent with comparable rotational and divergent imprints in highpass SF2s, which we also find (\S \ref{sec:highpass}).
Their lowpass statistics differ from ours: their large transverse-to-longitudinal SF2 ratio suggests rotationally dominated lowpass motions across diagnosed scales before de-correlation, whereas we find near-equipartition between rotational and divergent lowpass SF2s at $\rs\leq1$~km (\S \ref{sec:lowpass}).
Such differences may arise from data and methodology: 
they apply Eulerian filtering to gridded data, while we apply GLM-inspired Lagrangian filtering to surface-drifter observations. 
\cite{hypolite2026second} also report that interactions between surface waves and currents can produce small-scale motions with shallow SF2 slopes. Such effects are neglected by us and remain a caveat for our interpretations, especially as the drifters follow currents in the upper $1$~m. 

We find a robust power law of $\rs^{2/3}$ ($\kappa^{-5/3}$) below decorrelation  scales in highpass SF2s (\S \ref{sec:highpass}). In the Eulerian frame, a $\kappa^{-5/3}$ power law is often associated with an energy-cascade inertial range \citep{kolmogorov1941local}. Here, however, $\rs$ is defined in the GLM mean trajectory frame. As the coordinate transform from particle trajectory to mean trajectory is not necessarily volume-preserving \citep{McIntyre1988,Soward2010}, conservation laws and cascade arguments may differ between Eulerian and GLM coordinates \citep{buhler2014wave,Rocha2018}. 
Existing GLM conservation-law results for quasi-geostrophic dynamics \citep{Xie2015,Salmon2016} do not directly resolve the interpretation of highpass, unbalanced SF2 slopes.

Although we have argued for the mean trajectory frame, particle-frame artifacts need not always be large. 
In our analysis, the Helmholtz decomposition of highpass LASER data provides the clearest example of an unphysical outcome in the particle trajectory frame. For the Helmholtz decomposition of lowpass LASER SF2s and GLAD highpass/lowpass SF2s, the frame dependence is not conspicuous (not shown). 
In the wave-mean decomposition (\S \ref{sec:compromises}), the particle trajectory frame also does not differ perceptibly from the mean trajectory frame at separations above $500$~m. 
In simulations, the two frames may also be close when wave displacements are under-resolved on the model grid. 
That said, the choice of particle trajectory frame, rather than the mean trajectory frame, might be why Lagrangian filtering has sometimes been reported to generate unphysical outcomes. For example, in Fig.~5 of \cite{wang2023simple}, the balanced motion diagnosed from particle-trajectory Lagrangian filtering in a region with  strong internal waves is shown to contain visible wavy patterns.
In that example, strong internal waves may induce large wave displacements resolvable by the simulation,  making the artifact from particle trajectories visible.  As models reach finer resolutions, wave-induced displacements may become better resolved and the advantages of mean-trajectory attribution may become more evident.

The mean trajectory framework is implemented in recent grid-based Lagrangian filtering algorithms \citep{baker2025lagrangian,kafiabad2022grid}, which avoid costly particle tracking but require solving complex PDEs, and adaptations for realistic ocean models are ongoing. 
Implementations based on particle tracking are ongoing (Cai Maitland-Davies, private communication), but we are not aware of related published work.  \

As a focus of this work is to introduce GLM-inspired Lagrangian filtering to ocean observations, we intentionally study only one statistical metric: SF2s. The filtering can be applied to other drifter metrics. For example, lowpass third-order structure functions analogous to those of \citet{balwada2022direct}, computed from lowpass velocities and trajectories, can diagnose cross-scale kinetic energy transfer rates induced by mean flows.
The application of the GLM-inspired filtering to separate wave and mean flow impacts in other drifter statistics should be naturally motivated by different applications in future. 

\clearpage

\section*{Appendix A: Key subscripts} 
Key subscripts for variables in this paper are listed in Table \ref{subscripts}.
\begin{table} 
\caption{Key subscripts.} \label{subscripts}
\centering
\begin{tabular}{l c}
\hline
 $\mathrm{o}$ & Evaluated at unfiltered position, or belonging to unfiltered motions    \\
 $\mathrm{s}$  & Evaluated at lowpass filtered positions, or belonging to lowpass filtered motions   \\
 $\mathrm{f}$  & Belonging to highpass filtered motions   \\
 $\mathrm{L}$ & Induced by velocity differences along separation vectors   \\
 $\mathrm{T}$   & Induced by velocity differences across separation vectors    \\
 $\psi$  & Induced by rotational motions   \\
 $\phi$  & Induced by divergent motions   \\
 $W$ & Induced by linearized wave motions\\
\hline
\end{tabular}
\end{table}

\section*{Data Availability Statement}
Codes for structure function computations, diagnostics and plotting are available on Zenodo \citep{han_wang_2026_20496939}. The GLAD and LASER data products are available on \url{https://www.griidc.org/}.
\acknowledgments
Without implying their endorsement, we thank Oliver B\"{u}hler, Jacques Vanneste, J\"orn Callies,  Alexa Griesel, Julia Dr\"ager-Dietel, Xiaolong Yu and Leif Thomas for helpful discussions. 
This project originated at Courant Institute of Mathematical Sciences, USA, when all authors were working there. We acknowledge the use of ChatGPT for code generation and for stylistic edits. 
This paper is a contribution to the projects L3, M2, and  W2 of the Collaborative Research Centre TRR 181 ``Energy Transfers in Atmosphere and Ocean'' funded by the Deutsche Forschungsgemeinschaft (DFG, German Research Foundation) - Projektnummer 274762653, which supports HW. DB acknowledges support from NSF OCE- 2242109. 
JHX acknowledges support from the National Natural Science Foundation of China, grant numbers 12272006, 12472219, 42361144844 and 12588201, and from the Laoshan Laboratory under grant numbers LSKJ202202000, LSKJ202300100 and LSJKJ202400203.

\end{document}